\documentclass[aps,pre,twocolumn,superscriptaddress,amsmath,amssymb,longbibliography]{revtex4-2}

\usepackage{graphicx}
\usepackage{bm}
\usepackage{hyperref}

\newcommand{\vq}{\mathbf{q}}
\newcommand{\vR}{\mathbf{R}}
\newcommand{\vu}{\mathbf{u}}
\newcommand{\ve}{\mathbf{e}}

\graphicspath{{figures/}}

\begin{document}

\title{Phase behavior of two-dimensional star-polymer solutions with tunable attraction}

\author{Reyhaneh A. Farimani}
\email{reyhaneh.afghahi.farimani@univie.ac.at}
\affiliation{Faculty of Physics, University of Vienna, Boltzmanngasse 5, 1090 Vienna, Austria}

\author{Christos N. Likos}
\affiliation{Faculty of Physics, University of Vienna, Boltzmanngasse 5, 1090 Vienna, Austria}

\date{\today}

\begin{abstract}
We analyze the phase behavior of two-dimensional 
concentrated solutions of star polymers both in 
athermal solvents and in solvents of varying
quality employing recently-derived, accurately 
parametrized
pair interaction potential [R.\ A.\ Farimani
and C.\ N.\ Likos, Macromolecules \textbf{58}, 11827 (2025)]. By combining
an analytical, self-consistent harmonic approximation
with renormalization-group calculations, we draw the
phase boundaries of the solid phase, demonstrating that
it occupies a finite region in the density-temperature
phase diagram of star polymers, displaying re-entrant
melting at high densities. For athermal star polymers,
a lowest crystallization functionality $f_c \cong 12.7$
is established, below which solutions remain liquid
at all concentrations. The analytical results are
in quantitative agreement with computer simulations,
which establish a two-step, KTHNY-melting scenario 
via an intervening hexatic phase. Liquid-gas binodals
have been calculated by means of Gibbs Ensemble Monte Carlo simulations. We establish that for 
star functionality $f = 15$, the solid phase disappears
at supercritical temperatures but for $f = 20$ the extrapolated liquid-gas and liquid-solid phase boundaries cross, suggesting
a possible triple point.

\end{abstract}

\maketitle

\section{Introduction}
\label{sec:intro}
Soft-matter systems obey the same classical statistical mechanics as atomic ones,
with one crucial distinction: 
the Hamiltonian that governs the probability weights in their (coarse-grained) configuration space is an effective one. Colloidal particles inhabit the mesoscale,
and their mutual interactions emerge from tracing out a vast number of microscopic degrees of freedom 
— solvent, counterions, grafted chains — 
so that the effective potential encodes direct forces and medium-mediated ones alike. 
The price is that such potentials are constrained free energies rather than bare energies: 
state-dependent, in general many-body, and valid only within the conditions of the underlying coarse-graining~\cite{louis2002beware,vonferber2000triplet}.
The reward is the defining freedom of soft matter: 
because the interaction is emergent, it can be shaped by molecular design to an extent atomic matter never permits \cite{likos2001effective}. 
This freedom raises a question: how much of a system's macroscopic phase behavior is encoded in the shape of its effective pair interaction, and can it be extracted from the potential directly, rather than merely simulated?

The answer has two parts. The first is remarkably simple: 
for effective interaction potentials $V(r)$ that are
integrable and possess a Fourier transform 
$\hat V(q)$,
the topology of the phase diagram is determined by a single global property of the interaction potential, the sign structure of $\hat{V}(q)$ \cite{likos2001criterion}. If $\hat{V}(q)$ is positive definite and the potential is bounded, the system freezes and remelts upon compression. The solid occupies a closed region of the phase diagram, beyond which the fluid remains stable at all densities. If $\hat{V}(q)$ instead changes sign, the system forms cluster crystals \cite{likos2001criterion, prestipino2005gcm, pamies2009hertz}. This classification establishes the broad picture but leaves several questions open. How soft can the repulsion become before the solid region disappears? How does this region change when an additional attractive scale introduces fluid condensation? Answering these questions requires the second, more demanding step: locating the phase boundaries themselves.
Over the decades, researchers have investigated these boundaries using a range of methods: harmonic and self-consistent anharmonic descriptions of the solid~\cite{hooton1955anharmonicity,koehler1966scp,werthamer1970scp}, thermodynamic perturbation theories, classical density-functional theory, renormalization-group methods, and direct free-energy simulations \cite{lang2000gcm,prestipino2012elastic,wca1971,ramakrishnan1979dft,evans1979dft,likos2007why,vanteeffelen2006freezing,parola2008smooth,MendozaCoto2024,frenkel1984ladd,vega2008review}. Each has provided insight into particular systems, yet a general route from the bare interaction potential to the phase boundaries, without adjusting parameters along the way, has remained elusive.

Addressing these questions requires a physical system whose softness can be tuned through its molecular architecture. Star polymers, consisting of $f$ polymer chains anchored to a common core, provide a natural choice. Their effective pair potential is derived by integrating out the conformational degrees of freedom of their grafted, self-similar polymer structure rather than postulated.
Their pair repulsion is entropic: overlapping coronas restrict the conformations available to the arms. The interaction is logarithmic at short separations, reflecting these conformational constraints and the self-similar organization of the arms, and decays exponentially at larger separations. Scaling theory predicts an interaction amplitude proportional to $f^{d/(d-1)}$ in the large-$f$ limit for $d =2,3$, where $d$ is the spatial dimension. The arm number therefore controls the softness, from the chain limit toward increasingly strong repulsion at larger $f$. Yet the hard-sphere limit remains out of reach at any realizable arm number: the outermost blobs continue to fluctuate, leaving even a many-arm star with a hard core surrounded by a penetrable shell \cite{Likos1998,jusufi1999interaction,Watzlawek1999,Farimani2025}.

The phase behavior and glassy state of star polymers in three dimensions have been extensively investigated over the past three decades \cite{Watzlawek1999, 10.1021/ma801894x}. This study extends the analysis to a two-dimensional setting.
Two dimensions, however, are not merely a smaller arena: dimensionality acts on the problem twice.
First, dimensionality alters the interaction potential; under planar conformation, the interaction amplitude stiffens from $f^{3/2}$ to $f^2$ \cite{egorov2013adsorption,Farimani2025}.
Second, it reshapes the governing statistical mechanics, acting on the input.
Two-dimensional materials challenge the intuition developed from three-dimensional systems. At finite temperature, a two-dimensional crystal cannot maintain long-range translational order. Instead, the 2D solid exhibits quasi-long-range translational order and long-range orientational order, resulting in a finite shear modulus \cite{mermin1966wagner, hohenberg1967, mermin1968crystal}. Depending on the interaction potential, various melting pathways may occur \cite{strandburg1988,chui1983,bernard2011,kapfer2015,tsiok2020hertzian,thorneywork2017melting}. The most widely recognized mechanism is described by the Kosterlitz–Thouless–Halperin–Nelson–Young (KTHNY) theory, which involves defect-mediated melting through the successive unbinding of dislocations followed by disclinations \cite{kosterlitz1973, halperin1978, nelson1979, young1979}.
This defect-unbinding mechanism is closely related to that of the two-dimensional $XY$ model~\cite{berezinskii1971classical,kosterlitz1974xy}.

The window that persists across dimensions, however, emerges within an idealized description: ultrasoft model potentials are often purely repulsive, as is the entropic interaction between athermal stars\cite{Watzlawek1999, zu2016density}. Real soft matter systems generally also exhibit attraction, arising from van der Waals forces at the atomic scale or solvent-mediated effects at a coarser level of description. For star polymers, solvent quality provides a natural way to tune these interactions~\cite{likos1998theta,huissmann2009solvents}. As solvent quality worsens, monomers attract one another, and the effective star–star potential develops an attractive component. Here, the state dependence of the potential becomes useful: incorporating the thermal scale allows solvent quality to play the role of temperature, so that worsening solvent quality effectively cools the system by chemical means.
Attraction shifts the phase boundaries and, when sufficiently strong, induces gas–liquid coexistence~\cite{loverso2006attractions}. This raises the question of whether a gas–liquid–solid triple point exists. The Fourier-space criterion provides only a qualitative constraint: excessive attraction destroys the positivity of $\hat{V}(q)$, leading to cluster formation. Resolving the remaining effects of attraction requires the more demanding step of calculating the phase boundaries themselves.

Herein, we focus on thermal star polymers in two dimensions. This work discusses the phase behavior and disentangles how the potential's three features, softness, attraction, and arm span relative to the core size, affect the phase-diagram boundaries, using a theoretical approach and Monte Carlo simulation.
The potential input is in hand,
We have developed thermal stars' effective pair interaction through careful analysis in an earlier publication\cite{Farimani2025}. The theoretical model is also in hand; we incorporate an intelligent analytical method developed earlier by Mendoza-Coto \textit{et al.} \cite{MendozaCoto2024} by joining the simple harmonic scheme of solids with KTHNY renormalization-group flow so that a Young modulus computed from the bare potential decides the fate of dislocation unbinding; they have benchmarked on the Gaussian model. 
The star potential is a categorically harder passage, with a more complicated, attractive potential, and whether the route survives is tested by comparing against Monte-Carlo simulation results.

We map the phase boundary by Monte Carlo simulation in the athermal case in the density and arm number plane, and for the density-attraction strength plane for fixed arm number. We also sample the gas-liquid coexistence through Monte Carlo simulation in the Gibbs ensemble, and compare each melting line with the line predicted by the model developed by Mendoza-Coto \textit{et al.} \cite{MendozaCoto2024}.
On both boundaries, universality is our panacea. In the liquid-gas coexistence curve, the 2D Ising model fixes the exponents, and in the melting curve, the KTHNY flow shapes the melting separatrix. 

The rest of the manuscript is organized as follows: in Section \ref{sec:model}, we define the model and methods; in Section \ref{sec:mc}, we present the simulated phase diagrams, which are
compared with theoretical calculations in Section \ref{sec:theory}. In Section \ref{sec:knobs} we discuss the role of tail of the interaction and in Section
\ref{sec:conclusions} we present our summary and 
concluding remarks. Technical details on the theoretical
approximations, the critical point analysis for the 
liquid-gas transition and the phase detection of 
solid, hexatic, and liquid phases in terms of the
suitable correlation functions are shown in 
Appendixes \ref{app:scha}, \ref{app:gibbscrit}, and 
\ref{app:solidliquid}.
 \section{Model and methods}
\label{sec:model}

\subsection{Effective star--star interaction}
In our previous publication~\cite{Farimani2025},
the effective pair potential for both thermal and athermal star polymers within a quasi-two-dimensional framework has been computed.
The star polymers considered in this study possess $f$ arms, and solvent quality is modulated by the effective attraction between monomers, characterized by the parameter $\lambda$.
The value $\lambda = 0$ corresponds to the athermal limit, where the effective interaction is purely repulsive and entropic in origin.
An increase in $\lambda$ reduces solvent quality, resulting in a progressively more attractive effective interaction.
\label{sec:model:potential}

\begin{equation}
\frac{V(r)}{k_B T}
=
V_0(r; f, \alpha)
-\lambda f^2 A(\lambda)\exp[-\kappa(\lambda)\, r/\sigma],
\label{eq:veff}
\end{equation}
where the attractive contribution vanishes identically at $\lambda = 0$
and the repulsive core is expressed as:
\begin{equation}
V_0(r; f, \alpha)
=
\frac{2+9f^2}{24}
\begin{cases}
-\ln(r/\sigma)+\alpha, & r < \sigma, \\[6pt]
\alpha \exp\!\left[\dfrac{1 - r^2/\sigma^2}{2\alpha}\right], & r \ge \sigma.
\end{cases}
\label{eq:v0}
\end{equation}

Here, $\sigma$ represents the characteristic size of the star polymer (radius of gyration). The decay parameter $\alpha$ depends on both the functionality $f$
and the attraction strength $\lambda$, and it has been determined by matching the radial distribution
function of the effective fluid to results from monomer-resolved simulations,
to which the dilute regime is highly sensitive~\cite{Farimani2025}. 
Across the solvent-quality window examined in this work, it is smoothly
interpolated as
\begin{equation}
\alpha(f,\lambda) = \alpha_0(f)\, e^{-\lambda^2/(2s_f^2)},
\end{equation}
with
$(\alpha_0, s_f) = (0.872, 0.537)$ for functionality $f = 15$ and 
$(\alpha_0, s_f) = (0.832, 0.545)$
for functionality $f = 20$. In addition, the parameters entering the attractive 
tail of the potential are expressed as:
\begin{eqnarray}
A(\lambda) & = & 2.27 - 1.49\lambda,
\\
\kappa(\lambda) & = & 6.56 - 7.61\lambda.
\end{eqnarray}
The amplitude of the logarithmic core, $(2+9f^2)/24$, and attractive part $f^2$ are exact for all values of
$f$~\cite{Farimani2025}.
The effective pair potential and the interpolated
$\alpha(f,\lambda)$ are plotted for representative values of $f$ and
$\lambda$ in Section 1 of Appendix~\ref{app:scha}.

Two features of the effective interaction organize the phase behavior
studied below. First, since the coarse-graining absorbs the thermal
scale into the potential itself, $V(r)/k_B T$ is the complete
specification of the model and temperature is not an independent
control parameter. At $\lambda = 0$ the interaction is purely entropic,
so the athermal $(\rho, f)$ plane constitutes the \emph{complete} phase
diagram of the athermal-solvent system (Sec.~\ref{sec:mc:athermal}). For
$\lambda > 0$ the solvent quality takes over the role that temperature
plays in a conventional fluid --- increasing $\lambda$ strengthens the
attraction, analogous to cooling --- and phase diagrams are naturally
drawn in the $(\rho, \lambda)$ plane at fixed $f$. Second, the
effective-potential picture bounds the parameter window: we restrict
ourselves to $\lambda \le 0.6$, beyond which the Fourier transform
$\hat V(q)$ develops a negative region, signaling an instability toward
cluster or modulated phases outside the scope of this work.

\subsection{Monte Carlo simulations}
\label{sec:model:mc}

Bulk phase diagrams were generated using large-scale Metropolis Monte Carlo simulations.
These simulations were performed in the canonical ensemble, utilizing a C++ implementation parallelized with MPI.
For phase diagrams with $f \in \{15, 20\}$, the systems comprised $N = 2^{16} = 65,536$ point particles interacting via the effective potential. For other values of $f$, systems with $N = 2^{12} = 4096$ particles were examined.
Particles were placed in a rectangular simulation box with an aspect ratio
$L_x : L_y = \sqrt{3} : 2$, which is commensurate with triangular ordering.
Periodic boundary conditions were applied in both directions.
Simulations were initialized from fully random configurations. Additionally, a subset of runs was started from a fully triangular crystalline structure to validate the phase diagram.

Phases are identified using standard structural observables~\cite{nelson1979,keim2007frank}.
In the context of orientational-order, the bond-orientational
correlation function $g_6(r)$ is considered, which is
defined as:
\begin{equation}
g_6(r) = \left\langle \psi_6(\mathbf{r}_i) \psi_6^*(\mathbf{r}_j) \right\rangle,
\label{eq:c6}
\end{equation}
where $r = |\mathbf{r}_i - \mathbf{r}_j|$ is the separation
between particles $i$ and $j$ located at 
$\mathbf{r}_i$ and $\mathbf{r}_j$, respectively.
Here, the local bond-orientational
order parameter $\psi_6({\bf r}_j)$ is defined as:
\begin{equation}
\psi_6(\mathbf{r}_j) = \frac{1}{N_b(j)} \sum_{m=1}^{N_b(j)}  e^{6 {\mathrm{i}} \theta_{jm}},
\end{equation}
where the sum runs over the $N_b(j)$ nearest neighbors of particle $j$ determined by the Voronoi tesselation, 
and $\theta_{jm}$ is the angle between the bond connecting particles $j$ and $m$ and a fixed reference axis.
Moreover, the average $\langle \cdots \rangle$ is taken
over all particles and configurations.
To detect translational order, we considered the correlation 
function
\begin{equation}
c_G(r) = \left\langle e^{{\mathrm i} \mathbf{G} \cdot (\mathbf{r}_i - \mathbf{r}_j)}\right\rangle,
\label{eq:gg}
\end{equation}
where $\mathbf{G}$ is a reciprocal lattice vector of the triangular lattice and all other quantities 
and the average have the
same meaning as for the bond-orientational correlation
function defined in Eq.\ (\ref{eq:c6}) above.
The correlation functions $g_6(r)$ and
$c_G(r)$ are used to distinguish between solid, hexatic, and
fluid regions, which are classified based on the nature of their decay. In particular, in the fluid phase, both decay 
exponentially; in the hexatic phase, a power-law
decay results for $g_6(r) \sim r^{-\eta_6}$, with $\eta_6 < 1/4$,
whereas $c_G(r)$ decays exponentially. Finally, in the solid phase, $g_6(r)$ is essentially constant and
$c_G(r) \sim r^{-\eta_G}$ with $\eta_G < 1/3$.
Transition points are identified through
systematic scans of density at fixed $\lambda$ (
or fixed $f$ in the
athermal plane), combined with finite-size scaling analysis of the $\psi_6$ statistics.

Liquid--gas coexistence is sampled directly in the Gibbs
ensemble~\cite{Panagiotopoulos1987,Smit1989}: two coupled boxes at
fixed total particle number ($N = 4096$), total area, and common
$\lambda$ exchange area and particles through the standard three-move
scheme (particle displacements, area exchange, particle transfer),
thereby equilibrating pressure and chemical potential between the two
bulk phases without an explicit interface. The extraction of the coexistence
densities, and the critical-point analysis
are described together with the results in Sec.~\ref{sec:mc:lg}.

\subsection{Analytical framework: SCHA and renormalization group flow} 
\label{sec:model:theory}

\begin{figure}
    \centering
    \includegraphics[width=\linewidth]{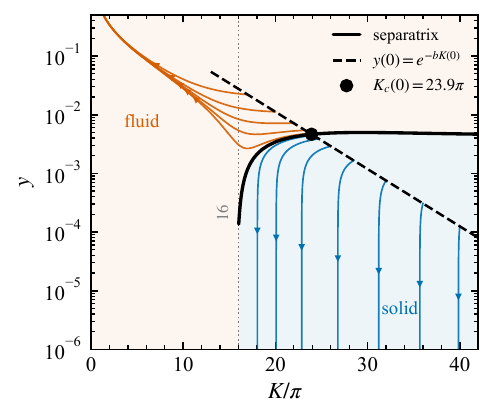}
    \caption{ RG flow trajectories according to 
    Eqs.\ (\ref{eq:rg:flowK}) and (\ref{eq:rg:flowy}) seeded on the locus, with arrows in the flow direction. Thick black: separatrix, asymptotic to the dotted 
    $K/\pi = 16$-line.
    The dashed line marks the microscopic locus $y(0) = e^{-bK(0)}$ and  the circle marks the crossing 
    of the two at $K_c(0) = 23.9\pi$, and $y(0) = 4.6 \times 10^{-3}$. Shading marks the two basins.}
    \label{fig:rgflow}
\end{figure}

The theoretical melting lines are computed from the pair potential
alone, with no fitted parameters, by combining a self-consistent
harmonic approximation (SCHA) for the elasticity of the solid~\cite{koehler1966scp,werthamer1970scp} with the
Kosterlitz--Thouless--Halperin--Nelson--Young (KTHNY)
renormalization group~\cite{nelson1979,young1979},
following the scheme of Ref.~\cite{MendozaCoto2024}. In the SCHA step
the triangular solid is described by the optimal Gaussian phonon
ensemble: the phonon propagator is determined self-consistently from
force constants that are themselves thermal averages of derivatives of
$V(r)$ over the phonon fluctuations, yielding fluctuation-dressed
elastic (Lam\'e) moduli $\mu(\rho,\lambda)$ and
$\lambda_L(\rho,\lambda)$ that soften on approach to melting. From
these the bare Young modulus
\begin{equation}
  K(0) = \frac{4\mu(\mu+\lambda_L)}{2\mu+\lambda_L}
  \label{eq:K0}
\end{equation}
is calculated, and it serves as the initial condition 
for the coupled KTHNY flow equations for the renormalized modulus $K(l)$ and the renormalized dislocation fugacity $y(l)$:
\begin{align}
\frac{dK^{-1}}{dl}
&=
\frac{3}{2}\pi y^2 e^{K/8\pi}
\left[
I_0\!\left(\frac{K}{8\pi}\right)
-
\frac{1}{2} I_1\!\left(\frac{K}{8\pi}\right)
\right],
\label{eq:rg:flowK}
\\
\frac{dy}{dl}
&=
\left(2-\frac{K}{8\pi}\right) y
+
2\pi y^2 e^{K/16\pi}
I_0\!\left(\frac{K}{8\pi}\right),
\label{eq:rg:flowy}
\end{align}
with $I_{0,1}(K/8\pi)$ are modified Bessel functions.

We assign to dislocation fugacity $y(l)$ the initial value
$y(0) = e^{-E_c/k_B T}$ set by the dislocation core
energy~\cite{nelson1979,sengupta2000elastic}. Following Ref.~\cite{MendozaCoto2024}, the core energy is taken
proportional to the bare modulus, $E_c/(k_BT) = b\,K(0)$
with $b = 0.072$; as these two are correlated, the melting
criterion then reduces to a critical value of the bare modulus alone,
$K_c(0) \simeq 23.9\pi$, the separatrix between flows that retain a
finite renormalized modulus [$K(l\to\infty) \ge 16\pi$, solid] and
those driven to the disordered fixed point (see Figure \ref{fig:rgflow})~\cite{vongrunberg2004elastic}. The locus of pairs $(\rho,\lambda_m(\rho))$
of the thermodynamic parameters yielding such Lam{\'e} 
moduli $(\mu,\lambda_L)$ that the bare young modulus
$K(0)$ takes the value $K_c(0)$
define the theoretical melting line $\lambda_m(\rho)$.

Because $\hat V(q) \sim 1/q^2$ at large $q$ for the star-polymer
potential, the SCHA self-consistency is solved in a real-space
formulation in which the momentum integrals are evaluated as
Gauss--Hermite quadratures of the Gaussian-smeared pair potential,
avoiding any momentum-space cutoff. The full self-consistent
equations as well as the numerical scheme are given in Appendix~\ref{app:scha}.

 \section{Phase behavior from simulation}
\label{sec:mc}

\subsection{Athermal limit: the $(\rho, f)$ phase diagram}
\label{sec:mc:athermal}
\begin{figure}
  \includegraphics[width=\columnwidth]{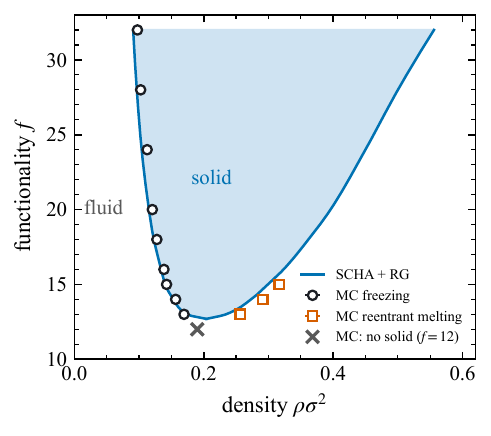}
  \caption{Athermal ($\lambda = 0$) phase diagram of the
  two-dimensional star-polymer solution in the $(\rho, f)$ plane.
  Symbols mark the Monte Carlo freezing (circles) and reentrant-melting
  (squares) densities, obtained by bisection to a bracket width
  $\Delta\rho\sigma^2 = 0.001$ ($N = 4096$; the symbol size exceeds the
  uncertainty). The solid line is the parameter-free SCHA+RG prediction, i.e.,
  the level set $K{(0)} = K_c{(0)}$ (shaded: solid); the two branches
  close at a critical functionality $f_c \simeq 12.7$, below which no
  crystal exists at any density. Consistently, at $f = 12$ the
  simulations find no solid even at the density of maximal predicted
  stability (cross).}
  \label{fig:lambda0_phase_diagram}
\end{figure}

In the athermal limit ($\lambda = 0$) the effective interaction is
purely repulsive, and the only relevant parameters are the density
$\rho$ and the arm number $f$; the resulting phase diagram in the
$(\rho, f)$ plane is shown in Fig.~\ref{fig:lambda0_phase_diagram}. 
To consider the density regime that the interaction is stay in the pair-wise additive region, we constraint ourselves to $\rho\sigma^2 \lesssim 0.35$, which our potential has proved to be valid \cite{Farimani2025}.
In contrast to the three-dimensional case, where star-polymer
solutions display a rich variety of crystalline structures\cite{likos2001effective}, the only
solid phase in two dimensions is the triangular lattice; consistent
with its being the densest packing in the plane.

Since the effective potential is positive definite for all $f$, the
system exhibits reentrant melting: the solid occupies a closed pocket
in the $(\rho, f)$ plane, which terminates at a critical functionality
$f_c \simeq 12.7$, below which no crystal exists at any density. In
three dimensions the corresponding threshold was found at
$f_c = 34$~\cite{likos2001effective}. The two numbers are not directly comparable, however, because the
core amplitude carries a different power of $f$ in the two
dimensionalities. Within the Daoud--Cotton picture the blob size at a
distance $r$ from the star centre follows from sharing the available
space among the $f$ arms: in three dimensions the surface
gives $\xi \sim r/\sqrt{f}$, whereas in two dimensions it gives $\xi \sim r/f$\cite{Daoud1982}. The number of
blobs per arm therefore grows as $\sqrt{f}$ rather than as $f$, and the
amplitude of the logarithmic core as $f^{3/2}$ rather than as $f^{2}$.
A given repulsion strength is thus reached at a much lower arm number
in two dimensions, and it is a direct consequence of the fact that steric repulsion is stronger as we decrease the dimensionality\cite{Farimani2025}.

\subsection{Liquid--gas coexistence}
\label{sec:mc:lg}
\begin{figure*}
  \includegraphics[width=\textwidth]{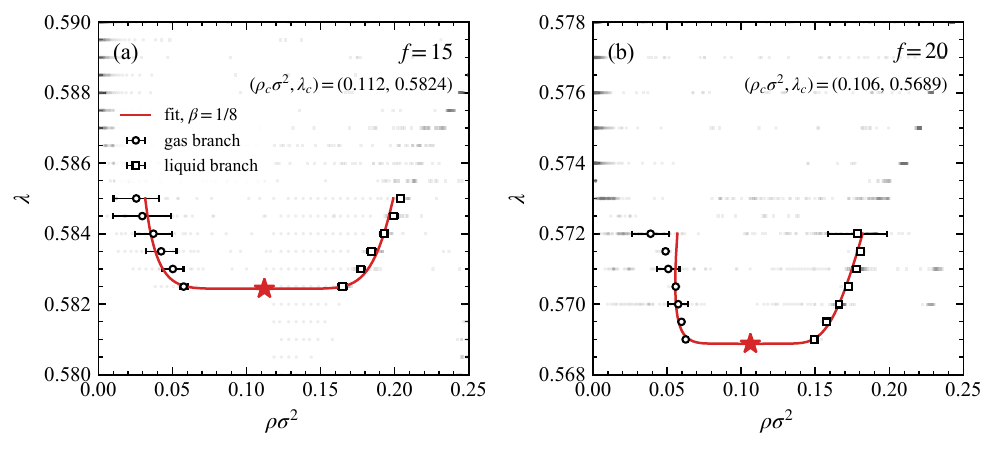}
  \caption{Liquid--gas binodals of the two-dimensional star-polymer
  solution, for
  (a)~$f = 15$ and (b)~$f = 20$. Gray points show the instantaneous
  densities of the two simulation boxes for every state point of the
  campaigns; open circles (squares) mark the gas (liquid) coexistence
  densities obtained as the per-$\lambda$ consensus of independent state
  points, after discarding runs whose imposed density lies at the edge
  of its own coexistence gap (lever-rule filter; see
  Appendix~\ref{app:gibbscrit}). Error bars give the spread across
  imposed densities. The red curve is a fit of the two-dimensional
  Ising form $\Delta\rho = B\,(\lambda/\lambda_c - 1)^{\beta}$ with
  $\beta = 1/8$ fixed and free amplitude $B$, together with a linear
  coexistence diameter; the star marks the resulting critical point,
  $(\rho_c, \lambda_c) = (0.112,\, 0.5824)$ for $f = 15$ and
  $(0.106,\, 0.5689)$ for $f = 20$. By construction, $\lambda_c$ lies
  below every state point that exhibits phase separation.}
  \label{fig:binodal}
\end{figure*}

As described in Sec.~\ref{sec:model:mc}, liquid--gas coexistence is
sampled in the Gibbs ensemble; Fig.~\ref{fig:binodal} shows the
resulting binodals for $f = 15$ and $f = 20$. At long wavelengths the
fluid is described by a coarse-grained scalar order-parameter field
$\phi(\mathbf{x}) \propto \rho(\mathbf{x}) - \rho_c$ with
$\mathbb{Z}_2$ symmetry, placing the transition in the two-dimensional
Ising universality class~\cite{wilding1992mixing}. Since the solvent quality controls the
attraction strength, $\lambda$ plays the role of an (inverse)
effective temperature (Sec.~\ref{sec:model:potential}), and assuming
the smooth mapping between the two to be linear near the critical
point, the coexistence curve follows~\cite{yang1952magnetization}
\begin{equation}
  \Delta\rho(\lambda) = \rho_L - \rho_g
  = B\,\bigl(\lambda/\lambda_c - 1\bigr)^{\beta},
  \qquad \beta = \tfrac{1}{8},
  \label{eq:isingfit}
\end{equation}
for $\lambda > \lambda_c$. The exponent is \emph{imposed} from
universality rather than fitted --- the sampled $\lambda$ window
cannot determine it, and only $\beta = 1/8$ reproduces the
characteristically flat top of a two-dimensional
binodal~\cite{Smit1991} --- while the amplitude $B$ is a free,
non-universal fit parameter. The coexistence diameter,
\begin{equation}
  \rho_{\mathrm{diam}}(\lambda)
  = \tfrac{1}{2}\bigl(\rho_L + \rho_g\bigr)
  \approx \rho_c + c\,(\lambda - \lambda_c),
  \label{eq:diameter}
\end{equation}
is approximated as linear near criticality and provides an estimate of the
critical density. Fitting
Eqs.~(\ref{eq:isingfit})--(\ref{eq:diameter}) to the consensus
coexistence densities yields the critical points
$(\rho_c, \lambda_c) = (0.112,\, 0.5824)$ for $f = 15$ and
$(0.106,\, 0.5689)$ for $f = 20$; the underlying fits of $\Delta\rho$
and $\rho_{\mathrm{diam}}$ are presented in
Appendix~\ref{app:gibbscrit}. It is worth noting that, since the
interfacial tension of these ultrasoft particles in two dimensions is
weak, we additionally examined the local-density distributions of
canonical ($NVT$) simulations to cross-check the coexistence results.

\subsection{Attractive stars: $(\rho, \lambda)$ diagrams at $f=20$ and $f=15$}
\label{sec:mc:lambda}
\begin{figure*}
  \includegraphics[width=\textwidth]{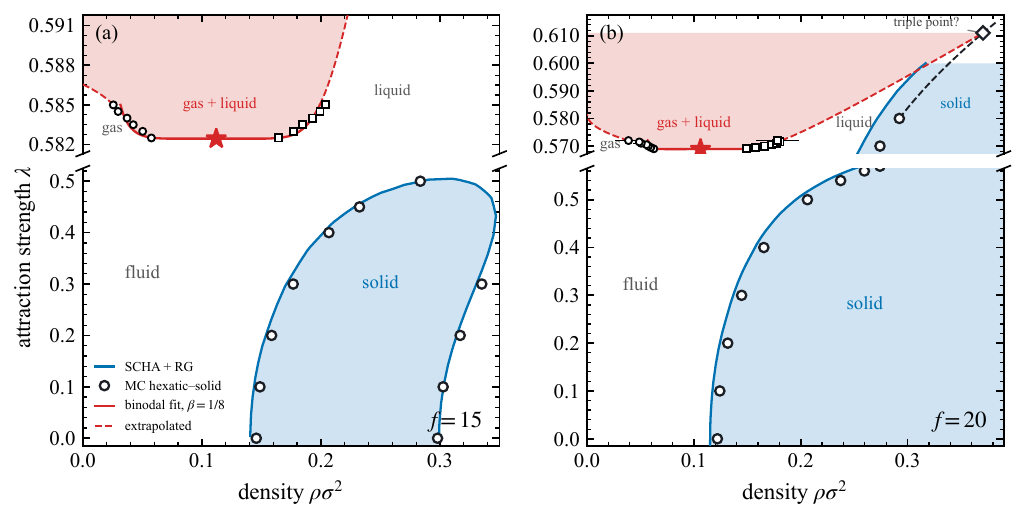}
  \caption{Phase diagrams of the star-polymer solution in the
  $(\rho, \lambda)$ plane, for (a)~$f = 15$ and (b)~$f = 20$. Circles
  mark the Monte Carlo hexatic--solid transition densities per solvent
  quality $\lambda$;
  the blue line and shaded region are the parameter-free SCHA+RG
  prediction of the solid pocket, which closes at
  $\lambda^{*} \simeq 0.505$ for $f = 15$, while for $f = 20$ it
  remains open at the $\lambda = 0.6$ cap (its reentrant branch lies
  beyond the plotted density range). The upper panels stretch the
  narrow solvent-quality window in which the fluid phase separates:
  open symbols show the Gibbs-ensemble coexistence densities, the red
  curves the two-dimensional Ising binodal fits [$\beta = 1/8$ imposed,
  Eq.~(\ref{eq:isingfit})] with critical points
  $(\rho_c, \lambda_c) = (0.112,\, 0.5824)$ and $(0.106,\, 0.5689)$
  (stars). Dashed lines extrapolate beyond the sampled windows: the gas
  branch follows the data trend and empties ($\rho_g \to 0$) at
  $\lambda \simeq 0.587$~(a) and $\simeq 0.582$~(b), the liquid branch
  follows the fit form, and in~(b) the continued MC hexatic--solid
  line meets the extrapolated liquid branch at a possible
  gas--liquid--solid triple point near
  $(\rho\sigma^2, \lambda) \simeq (0.37,\, 0.61)$ (diamond).}
  \label{fig:phase_diagram_f15_f20}
\end{figure*}

Away from the athermal limit, the solvent quality $\lambda$ acts as
the second control parameter; Fig.~\ref{fig:phase_diagram_f15_f20}
shows the resulting $(\rho, \lambda)$ phase diagrams for $f = 15$ and
$f = 20$. For the melting points reported here the particle number was
increased to $N = 2^{16}$ to reduce finite-size effects.

Complementing the liquid--gas coexistence windows and critical points
of Sec.~\ref{sec:mc:lg}, we determined the melting densities across
the full range of solvent qualities. With increasing $\lambda$ ---
i.e., increasing effective attraction --- the freezing density rises,
as expected: at fixed density, the attraction destabilizes the
crystal.

For $f = 15$ no solid phase is found for $\lambda \gtrsim 0.5$: the
solid pocket closes at $\lambda^{*} \simeq 0.5$, well below the
critical solvent quality $\lambda_c = 0.5824$. The liquid--gas
coexistence window therefore opens in a purely fluid regime, and no
triple point exists. For $f = 20$, by contrast, the solid pocket
remains open above $\lambda = 0.6$, and the continued melting
line meets the extrapolated liquid branch at a possible
gas--liquid--solid triple point near
$(\rho\sigma^2, \lambda) \simeq (0.37,\, 0.61)$. Increasing the arm
number thus stabilizes the solid phase to higher solvent qualities,
suggesting a possible triple point beyond the directly studied range of
the effective potential. The analytical melting boundaries included in
Fig.~\ref{fig:phase_diagram_f15_f20} are introduced and compared with
these data in Sec.~\ref{sec:theory}. \section{Parameter-free theory versus simulation}
\label{sec:theory}

The blue curves in Figs.~\ref{fig:lambda0_phase_diagram}
and~\ref{fig:phase_diagram_f15_f20} are the melting lines predicted by
the analytical framework of Sec.~\ref{sec:model:theory}, following
the scheme of Mendoza-Coto \emph{et al.}~\cite{MendozaCoto2024}: the
SCHA supplies the fluctuation-dressed elastic moduli
$\mu(\rho,\lambda)$ and $\lambda_L(\rho,\lambda)$, and melting lies
on the locus where the bare Young modulus of Eq.~(\ref{eq:K0})
crosses the critical value $K_c(0) \simeq 23.9\pi$. Every point of
the theoretical boundaries in the $(\rho, f)$ and $(\rho, \lambda)$
planes follows from this single condition, with the effective pair
potential of Eqs.~(\ref{eq:veff})--(\ref{eq:v0}) as the only input
and no adjustable parameters. The scheme was originally benchmarked
on the Gaussian core model; applying it here to a considerably more
complex interaction --- a logarithmically divergent core with a
slowly decaying Fourier tail, a tunable attraction, and two control
parameters --- requires only a real-space reformulation of the
numerical solution (Appendix~\ref{app:scha}).

In the athermal plane (Fig.~\ref{fig:lambda0_phase_diagram}), the
freezing branch is reproduced quantitatively across the entire range
of functionalities studied: the predicted freezing densities lie
within $0.004$--$0.009$ in $\rho\sigma^2$ of the Monte Carlo values
for all $f = 13$--$32$, a relative deviation of a few percent with no
systematic degradation up to the largest arm numbers. The theory also
reproduces the existence and location of the lower critical
functionality: the predicted closure of the solid pocket at
$f_c \simeq 12.7$ falls inside the Monte Carlo bracket --- a solid
exists at $f = 13$ but not at $f = 12$, even at the density of
maximal predicted stability. The reentrant branch is captured
qualitatively but is systematically underestimated: for
$f = 13$--$15$ the measured reentrant-melting densities exceed the
predicted ones by $\simeq 0.02$ in $\rho\sigma^2$, i.e.\ the actual
crystal survives to somewhat higher densities than the harmonic
description allows.

The same pattern carries over to the attractive plane
(Fig.~\ref{fig:phase_diagram_f15_f20}). For both functionalities the
theoretical freezing line tracks the Monte Carlo points to
$\sim 0.005$ in $\rho\sigma^2$ at small $\lambda$, with the deviation
growing to $\sim 0.02$ only in the strongly attractive regime
$\lambda \gtrsim 0.5$. Most notably, the topology of the two diagrams
is predicted correctly: for $f = 15$ the theory closes the solid
pocket at $\lambda^{*} \simeq 0.505$, in near-perfect agreement with
the simulations, which find no solid for $\lambda \gtrsim 0.5$, while
for $f = 20$ the pocket is correctly predicted to remain open over
the entire accessible solvent-quality range.

This level of agreement from a parameter-free calculation
demonstrates the predictive power of the approach of
Ref.~\cite{MendozaCoto2024} well beyond the system it was tested for. The KTHNY criterion is crossed while the system is still a
well-ordered solid --- at the boundary the bare modulus is
still above the universal value $16\pi$, and the defect-pair density
is small --- so the Gaussian-phonon (SCHA) description of the input
elasticity is used only in a regime where it is accurate. In
addition, near the freezing branch $K(0)$ is a steep function of
density, so even a sizable relative error in the criterion translates
into a small shift of the predicted boundary. The two places where
the theory is least accurate are precisely those where these
conditions degrade: on the reentrant side and at large $\lambda$ where melting density is higher, and therefore anharmonic corrections grow; also, the dislocation core energy
--- taken here with the core scale $d = \sqrt{3}\,a$ inherited from
Ref.~\cite{MendozaCoto2024} --- is the least controlled ingredient of
the calculation. We also note that the theoretical line marks the
dislocation-unbinding (solid--hexatic) instability, whereas the
simulations locate the loss of translational order; at the resolution
of Figs.~\ref{fig:lambda0_phase_diagram}
and~\ref{fig:phase_diagram_f15_f20} the width of the conventional
hexatic window is smaller than the symbol size
(Sec.~\ref{sec:model:mc}), so the two notions of the boundary
coincide here.

 \section{The tail effect}
\label{sec:knobs}
While the core of the star--star potential follows from self-similarity, its tail is more complex and can be substantially modified by chemical details, such as electrostatic interactions and additional functional groups. To examine how changes in the tail affect the phase diagram, we multiplied $\alpha(f)$ in our potential by a uniform scaling factor and calculated the resulting phase diagrams.

\begin{figure*}
    \centering
    \includegraphics[width=\linewidth]{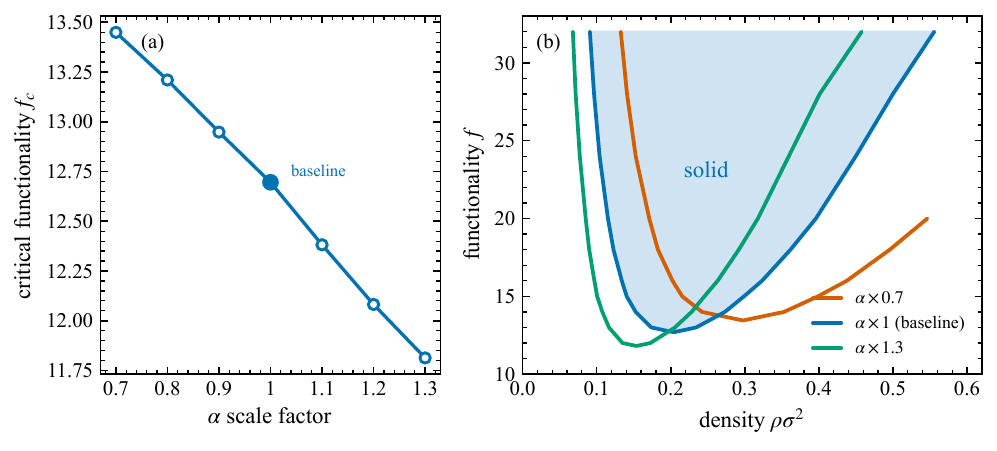}
    \caption{Effect of the soft repulsive tail on athermal melting.
    (a) Critical functionality $f_c$ as a function of the uniform scaling factor applied to $\alpha(f)$.
    (b) Region of solid stability for different scaling factors.}
    \label{fig:tail}
\end{figure*}

The results are shown in Fig.~\ref{fig:tail}. A narrower tail, obtained by multiplying $\alpha(f)$ by $0.7$, reduces the interaction range and shifts both the low-density and reentrant melting boundaries to higher densities. Conversely, a broader tail, obtained by multiplying $\alpha(f)$ by $1.3$, increases the interaction range and shifts both boundaries to lower densities. The critical functionality $f_c$ decreases strictly monotonically with the scaling factor applied to $\alpha(f)$. These results highlight the combined role of the range, amplitude, and shape of the potential in determining the phase diagram, showing that even small variations in the tail can influence the phase behavior~\cite{bos2019tails,zhu2019layer}. \section{Conclusions}
\label{sec:conclusions}

We have determined the phase behavior of two-dimensional star-polymer solutions as a function of arm number $f$ and solvent quality $\lambda$, combining large-scale Monte Carlo simulations with a melting theory that couples the self-consistent harmonic approximation to the KTHNY renormalization-group flow. Both approaches use the effective pair potential of Ref.~\cite{Farimani2025} as their sole input. Its logarithmic core, slowly decaying Gaussian tail, and solvent-tunable attraction provide a more demanding test of the theory than the Gaussian core model for which it was originally developed~\cite{MendozaCoto2024}.

Under athermal solvent conditions, the triangular crystal occupies a closed, reentrant pocket in the $(\rho,f)$ plane. This pocket terminates at a lower critical functionality $f_c \simeq 12.7$: simulations find a stable solid at $f=13$, but none at $f=12$, even at the density of maximal stability. This threshold is substantially lower than its three-dimensional counterpart, $f_c=34$, reflecting the stronger dependence of the logarithmic-core amplitude on functionality, which scales as $f^2$ in two dimensions rather than $f^{3/2}$.

Worsening solvent quality introduces attraction, shifting the freezing boundary to higher densities and shrinking the solid pocket. At sufficiently poor solvent quality, the fluid separates into coexisting gas and liquid phases. Gibbs-ensemble simulations, analyzed assuming two-dimensional Ising criticality, locate the critical points at $(\rho_c\sigma^2,\lambda_c) = (0.112, 0.5824)$ for $f=15$ and $(0.106, 0.5689)$ for $f=20$. The relative positions of the melting and coexistence boundaries depend on functionality. For $f=15$, the solid pocket closes at $\lambda^* \simeq 0.5$, before condensation begins. Gas--liquid coexistence therefore occurs entirely within the fluid regime, and no triple point exists. For $f=20$, the solid persists up to the limit of the accessible solvent-quality range.

The SCHA+RG theory reproduces the solid boundaries without fitting parameters to the phase-behavior data. Two features explain its accuracy. First, the KTHNY criterion is reached while the solid remains well ordered, where a Gaussian description of phonon fluctuations accurately captures its elasticity. Second, near the freezing branch, the bare Young modulus varies steeply with density, so errors in the melting criterion produce only small shifts in the predicted boundary. Deviations become more pronounced where these conditions weaken and anharmonic effects grow, particularly near the reentrant melting branch for $f=15$ and at high $\lambda$ for $f=20$.

These results address the central question posed at the outset: for the systems considered here, phase boundaries can be predicted directly from the effective pair potential, even in the presence of a logarithmically soft core and solvent-mediated attraction. The remaining errors are small and can be traced to identifiable physical limitations of the theory.

The remaining questions concern the order of the transitions, the melting pathway, and the possible existence and width of a hexatic phase. A companion paper will examine these questions alongside defect statistics, extending the present connection between the pair potential and the phase diagram to the mechanisms of two-dimensional melting. 
\begin{acknowledgments}
 We acknowledge support from the European Union (Horizon-MSCA-Doctoral  Networks) through the project QLUSTER (HORIZON-MSCA-2021-DN-01-GA101072964).
The calculations were performed at the Austrian Scientific Cluster (ASC).

\end{acknowledgments}

\appendix

\section{SCHA formulation and real-space solver}
\label{app:scha}

\subsection{Effective pair potential}

\begin{figure*}
  \includegraphics[width=\textwidth]{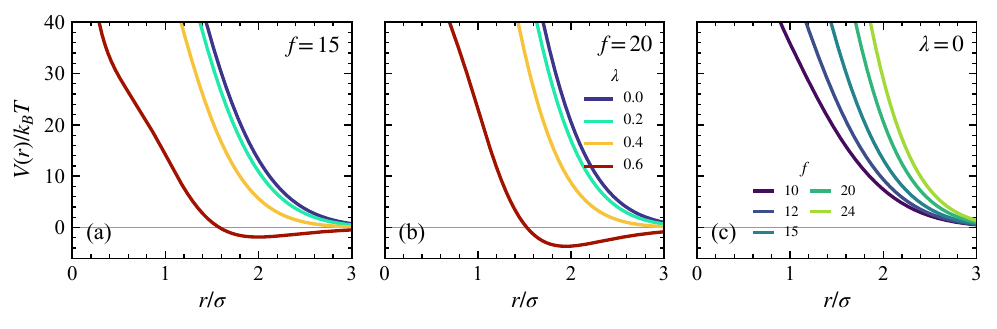}
  \caption{Effective star--star pair potential,
  Eqs.~(\ref{eq:veff})--(\ref{eq:v0}). (a),(b)~$f = 15$ and $f = 20$
  at solvent qualities $\lambda = 0$, $0.2$, $0.4$, and $0.6$: the
  ultrasoft core diverges only logarithmically at overlap, and
  increasing $\lambda$ strengthens the exponential attraction, which
  develops a well of a few $k_BT$ (region below the gray zero line)
  around $r \approx 2\sigma$ at the largest $\lambda$ studied.
  (c)~Athermal limit $\lambda = 0$ across functionality,
  $f = 10$ to $f = 24$: the purely repulsive, entropic potential stiffens
  with $f$ through the exact core amplitude $(2+9f^2)/24$.}
  \label{fig:app:potential}
\end{figure*}

We calculated the effective pair potential between two star polymers in the slit geometry and verified it by comparing the pair-distribution function of the monomer-resolved and analytical forms for different star polymers (functionality and solvent quality) in reference \cite{Farimani2025}. We present the resulting analytical function in Fig. \ref{fig:app:potential}.

\begin{figure*}
  \includegraphics[width=\textwidth]{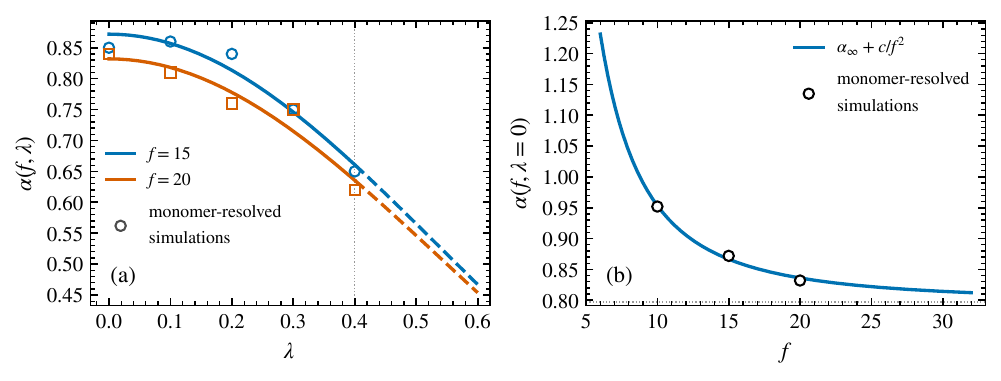}
  \caption{Tail parameter $\alpha$ of Eq.~(\ref{eq:v0}).
  (a)~Solvent-quality dependence $\alpha(f,\lambda)$ for $f = 15$ and
  $f = 20$, from the Gaussian interpolation
  $\alpha = \alpha_0(f)\, e^{-\lambda^2/(2s_f^2)}$ with the parameters
  quoted in Sec.~\ref{sec:model:potential}. Solid curves lie within
  the window $\lambda \le 0.4$ of the underlying monomer-resolved
  fits~\cite{Farimani2025}; dashed curves show the extrapolation of
  the same formula used for $0.4 < \lambda \le 0.6$ (dotted line:
  fit-window boundary). (b)~Athermal values $\alpha(f, \lambda = 0)$:
  open circles are the tabulated amplitudes $A(f)$ for
  $f = 10$, $15$, $20$~\cite{Farimani2025}; the curve is the
  least-squares law $\alpha_\infty + c/f^2$
  ($\alpha_\infty = 0.797$, $c = 15.65$, dotted line at
  $\alpha_\infty$) used for all other functionalities in the athermal
  $(\rho, f)$ phase diagram.}
  \label{fig:app:alpha}
\end{figure*}

\subsection{Variational free energy}

We describe the solid by displacements $\vu(\vR_j)$ of the particles
about the sites $\vR_j$ of a triangular lattice of spacing
$a(\rho) = [2/(\sqrt{3}\rho)]^{1/2}$, and follow the self-consistent
harmonic scheme of Ref.~\cite{MendozaCoto2024}. The trial ensemble is
generated by a Gaussian Hamiltonian diagonal in the Fourier modes
$\hat{\vu}(\vq)$ of the displacement field,
\begin{widetext}
\begin{equation}
  H_0 = N\epsilon_0(\rho)
  + \frac{1}{2\rho}\int_{\mathrm{BZ}}\frac{{\rm d}^2q}{(2\pi)^2}
  \Bigl[ f_1(\vq)\,|\hat u_\parallel(\vq)|^2
       + f_2(\vq)\,|\hat u_\perp(\vq)|^2 \Bigr],
  \label{eq:app:H0}
\end{equation}
\end{widetext}
where $\epsilon_0(\rho)$ is the static lattice energy per particle,
$\hat u_{\parallel,\perp}$ are the components of $\hat{\vu}$ parallel
and perpendicular to $\vq$, and $f_1(\vq)$, $f_2(\vq)$ are, at this
stage, \emph{arbitrary} longitudinal and transverse dispersion
functions on the first Brillouin zone (BZ). They are fixed by the
Gibbs--Bogoliubov inequality, which bounds the true free energy from
above by
\begin{equation}
  F \le F_{\mathrm{var}}[f_1, f_2]
  = F_0 + \langle H - H_0 \rangle_0,
  \label{eq:app:GB}
\end{equation}
where $\langle\cdot\rangle_0$ and $F_0$ are the average and free
energy of the trial ensemble~(\ref{eq:app:H0}); the optimal
dispersions minimize $F_{\mathrm{var}}$. Because $H_0$ is Gaussian,
the only nontrivial term is the average of the pair interaction,
\begin{equation}
  \langle H \rangle_0
  = \frac{N}{2}\sum_{\vR \neq 0}
  \bigl\langle V\!\bigl(\vR + \delta\vu(\vR)\bigr) \bigr\rangle_0,
  \label{eq:app:Vavg}
\end{equation}
taken over the relative displacement of the bond,
$\delta\vu(\vR) = \vu(\vR) - \vu(\mathbf{0})$.

\subsection{Self-consistent equations}

Within the harmonic ensemble $\delta\vu(\vR)$ is a zero-mean Gaussian,
so its statistics are carried entirely by its characteristic function,
\begin{equation}
  \bigl\langle e^{i \vq\cdot\delta\vu(\vR)} \bigr\rangle_0
  = e^{-\vq^{\mathsf T} M(\vR)\, \vq},
  \label{eq:app:gq}
\end{equation}
with covariance $\langle \delta u_a(\vR)\, \delta u_b(\vR)\rangle_0 = 2M_{ab}(\vR)$
($a,b \in \{x,y\}$),
obtained from Eq.~(\ref{eq:app:H0}) by equipartition. The matrix
$M(\vR)$ is given by the expression:
\begin{widetext}
\begin{equation}
  M(\vR) = \frac{1}{\beta\rho} \int_{\mathrm{BZ}}
  \frac{{\rm d}^2q}{(2\pi)^2}\, \bigl[1 - \cos(\vq\cdot\vR)\bigr]
  \left[ \frac{\vq\otimes\vq}{q^2 f_1(\vq)}
       + \frac{\vq_\perp\otimes\vq_\perp}{q^2 f_2(\vq)} \right],
  \label{eq:app:M}
\end{equation}
\end{widetext}
where $\vq = (q_x, q_y)$ and  
$\vq_\perp = (-q_y, q_x)$; the $[1-\cos(\vq\cdot\vR)]$ factor is the
structure factor of the \emph{relative} displacement of the two
sites, and the two terms are the longitudinal and transverse
contributions weighted by their inverse stiffnesses $1/f_{1,2}$.
Inserting Eq.~(\ref{eq:app:gq}) into the interaction average and
carrying out the Gaussian average of $V$ over $\delta\vu$ makes
$\langle H\rangle_0$ an explicit functional of $f_1, f_2$; imposing
stationarity of the bound~(\ref{eq:app:GB}),
$\delta F_{\mathrm{var}}/\delta f_i(\vq) = 0$, yields the
self-consistency condition that the trial force constants equal the
thermal averages of the true ones over the ensemble's own
fluctuations,
\begin{equation}
  f_i(\vq) = \sum_{\vR \neq 0} \bigl[1 - \cos(\vq\cdot\vR)\bigr]\,
  \hat{\ve}_i(\vq) \cdot \Phi(\vR) \cdot \hat{\ve}_i(\vq),
  \qquad i = 1, 2,
  \label{eq:app:selfcons}
\end{equation}
where $\hat{\ve}_1 = \hat{\vq}$, $\hat{\ve}_2 = \hat{\vq}_\perp$, and
$\Phi(\vR)$ is the phonon-smeared Hessian of the pair potential,
\begin{equation}
  \Phi_{ab}(\vR)
  = \bigl\langle \partial_a \partial_b V(\vR + \vu) \bigr\rangle_{
    \vu \sim \mathcal{N}(0,\, 2M(\vR))}.
  \label{eq:app:phi}
\end{equation}
Equations~(\ref{eq:app:M})--(\ref{eq:app:phi}) close on themselves
[they are the unreduced Eq.~(15) of Ref.~\cite{MendozaCoto2024}]: the
dispersions set the smearing widths $M(\vR)$, and the smeared
potential regenerates the dispersions. No small-$\vq$ expansion or
quadratic-dispersion ansatz is made; the equations are solved on the
full Brillouin zone. The physical picture is transparent: phonon
fluctuations blur each lattice bond by the Gaussian of
Eq.~(\ref{eq:app:gq}), and the elastic response is that of the
blurred potential --- for ultrasoft particles this dressing is strong
and is what drives the melting-line shapes of Sec.~\ref{sec:theory}.

The dressed elastic moduli follow from the acoustic (long-wavelength)
limit of the converged dispersions,
\begin{equation}
  2\mu + \lambda_L = \lim_{q\to 0}\frac{f_1(\vq)}{q^2},
  \qquad
  \mu = \lim_{q\to 0}\frac{f_2(\vq)}{q^2},
  \label{eq:app:lame}
\end{equation}
which are the phonon-dressed Lam\'e coefficients entering the Young
modulus $K(0)$ of Eq.~(\ref{eq:K0}) and hence the melting criterion.
In the $\beta \to \infty$ limit the smearing vanishes,
$M(\vR) \to 0$ and $\Phi(\vR) \to \partial\partial V(\vR)$, and
Eqs.~(\ref{eq:app:selfcons})--(\ref{eq:app:lame}) reduce to the bare
$T = 0$ harmonic lattice sums; for the star polymer $\beta = 1$ is
fixed, so the moduli retain their fluctuation dressing at every state
point and the bare sums serve only as a low-temperature check of the
solver.

The key computational step is the observation, exact for any Gaussian
ensemble, that the correlator exponent in Eq.~(\ref{eq:app:gq}) is a
$\vq$-independent quadratic form: $M(\vR)$ is computed once per
lattice vector and iteration, after which the momentum integral
implicit in Eq.~(\ref{eq:app:phi}) collapses to three scalar moments
per $\vR$. This reduces a nominally nested four-dimensional
integration to a cost linear in the number of lattice vectors and
grid points.

\subsection{Real-space (Gauss--Hermite) evaluation}

Written in momentum space, Eq.~(\ref{eq:app:phi}) reads
\begin{equation}
\Phi_{ab}(\vR) = -\int \! \frac{{\rm d}^2q}{(2\pi)^2}\, e^{i\vq\cdot\vR}
\hat V(q)\, q_a q_b\, e^{-\vq^{\mathsf T} M(\vR) \vq}, 
\end{equation}
which is an integral over
the whole plane. For the star-polymer potential this form is
numerically unusable: the logarithmically divergent core of
Eq.~(\ref{eq:v0}) gives $\hat V(q) \sim 1/q^2$ at large $q$, and no
practical momentum cutoff converges. We therefore evaluate
Eq.~(\ref{eq:app:phi}) directly in real space: the average of the
Hessian over $\vu \sim \mathcal{N}(0, 2M)$ is computed by
tensor-product Gauss--Hermite quadrature in the
eigenbasis of the $2\times 2$ matrix $M(\vR)$, using the closed-form
derivatives of the isotropic potential. This representation is
mathematically identical to the momentum integral --- no Fourier
transform of $V$ is ever taken --- and its quadrature error is
spectrally small.

\subsection{Iteration, moduli, and validation}

The fixed point is reached by Picard iteration with linear mixing,
initialized from the $T = 0$ (bare) Lam\'e coefficients, and is
declared converged when the relative sup-norm change of $f_1, f_2$
falls below $10^{-7}$; the production calculations use a
Brillouin-zone grid of $n = 24$--$32$ points per reciprocal basis
vector. Loss of a positive solution ($f_i > 0$) marks the SCHA
spinodal --- the mean-field limit of stability of the variational
solid --- which lies well beyond the RG melting line everywhere in
this work. The dressed Lam\'e moduli entering Eq.~(\ref{eq:K0}) are
read off from the acoustic limits $f_1(\vq) \to (\lambda_L + 2\mu)
q^2$ and $f_2(\vq) \to \mu q^2$, extrapolated to $q \to 0$ by a
linear fit of $f_i/q^2$ in $q^2$ over the smallest-$|\vq|$ grid
points.

\begin{figure*}
  \includegraphics[width=\textwidth]{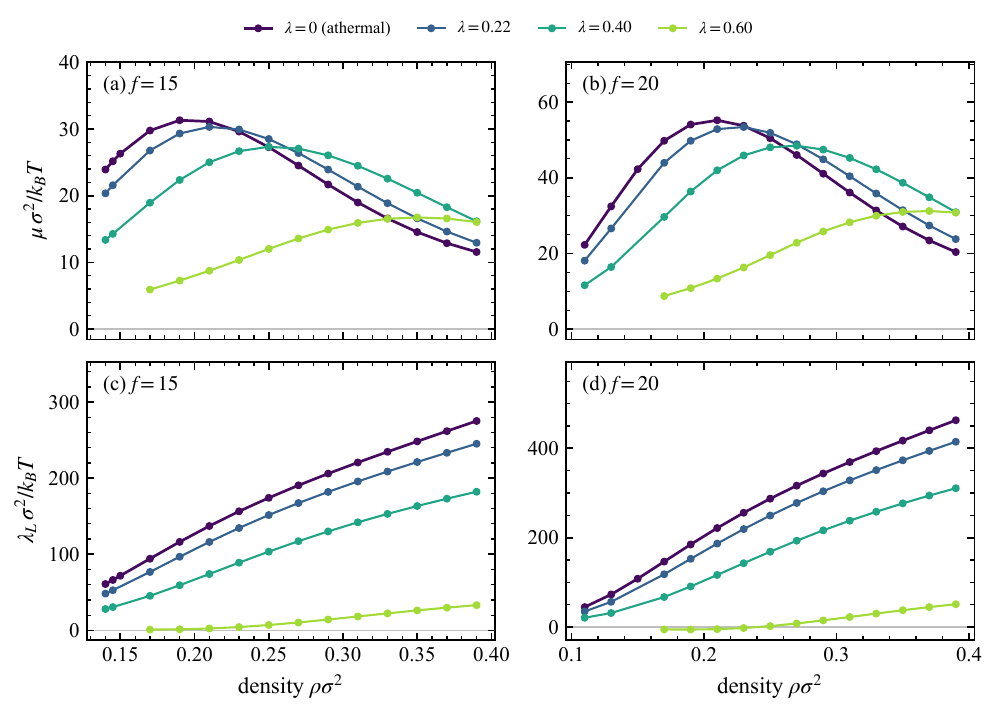}
  \caption{SCHA-dressed Lam\'e coefficients of the star-polymer solid:
  shear modulus $\mu$ (top row) and $\lambda_L$ (bottom row), for
  $f = 15$ [(a),~(c)] and $f = 20$ [(b),~(d)], in the athermal limit
  $\lambda = 0$ and at the thermal solvent qualities
  $\lambda \simeq 0.22$, $0.40$, and $0.60$ (symbols: converged
  full-Brillouin-zone solutions of
  Eqs.~(\ref{eq:app:M})--(\ref{eq:app:phi}); units $k_B T/\sigma^2$).
  The shear modulus is non-monotonic in density, rising as the
  crystal forms and softening again as the ultrasoft cores overlap and
  the lattice approaches reentrant melting, whereas $\lambda_L$
  increases monotonically. Increasing the attraction softens the solid
  dramatically: at $\lambda = 0.6$ both moduli are reduced by up to an
  order of magnitude relative to the athermal limit, and $\lambda_L$
  even turns slightly negative at low density for $f = 20$, while the
  bulk combination $\mu + \lambda_L$ and the Young modulus of
  Eq.~(\ref{eq:K0}) remain positive throughout the solid. Each curve
  spans the densities at which the self-consistency admits a solid
  solution at that $\lambda$.}
  \label{fig:app:lame}
\end{figure*}

Figure~\ref{fig:app:lame} shows the resulting dressed Lam\'e
coefficients $\mu(\rho, \lambda)$ and $\lambda_L(\rho, \lambda)$ for
both functionalities, in the athermal limit and at three thermal
solvent qualities.

\section{Critical-point analysis of the liquid--gas transition}
\label{app:gibbscrit}

Each Gibbs-ensemble state point is
defined by the solvent quality $\lambda$ and the imposed overall
density $\rho$. From the instantaneous densities of the two boxes we
form the liquid-like and gas-like branches,
$\rho_L(t) = \max[\rho_A(t), \rho_B(t)]$ and
$\rho_g(t) = \min[\rho_A(t), \rho_B(t)]$, and their separation
$\Delta\rho(t)$. Well-demixed configurations are selected by a
threshold $\tau$ taken as the mode of the smoothed histogram of
$\Delta\rho(t) > 0$; the coexistence densities of the state point are
the mean branch densities over the frames with $\Delta\rho(t) > \tau$,
with the branch widths as their uncertainties. A state point is
accepted as phase separated only if the resulting gap exceeds both a
minimum value and twice the combined branch widths, and the pooled
box-density histogram shows two clearly resolved peaks; the latter
condition rejects near-critical runs in which the boxes exchange
identity or demix only partially.

At fixed $\lambda$ several
independent state points with different imposed $\rho$ sample the same
tie line, and the lever rule locates each within its own gap,
$x = (\rho - \rho_g)/(\rho_L - \rho_g)$. State points with
$x < 0.15$ or $x > 0.85$ are excluded: there one box holds almost the
entire system and its partner is a degenerate remnant, so the inferred
``coexistence'' densities merely reproduce the imposed bulk fluid.
The surviving state points are aggregated per $\lambda$ into consensus
branches (median, with the interquartile range as uncertainty), shown
as the open symbols in Fig.~\ref{fig:binodal}. The enhanced scatter of
the gas branch at larger $\lambda$ has a physical origin --- the gas
becomes so dilute that a single box barely holds it --- and is
therefore down-weighted through its error bars rather than discarded.
\begin{figure*}
  \includegraphics[width=\textwidth]{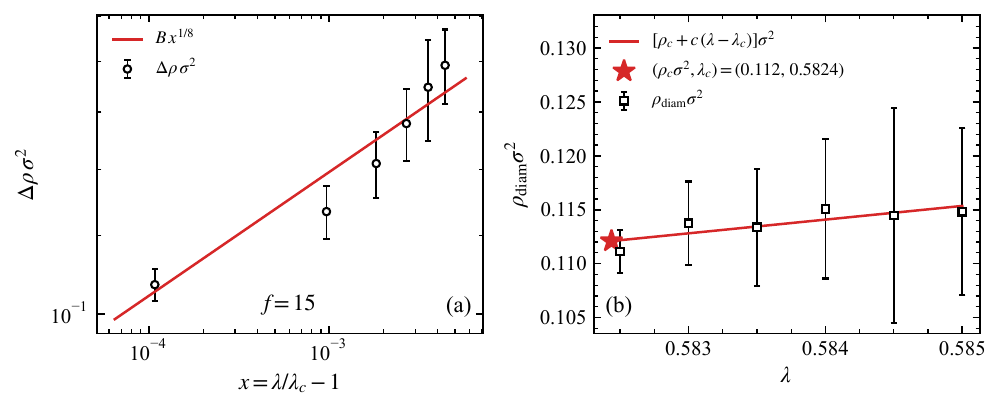}
  \caption{Critical scaling of the liquid--gas transition at $f = 15$.
  (a)~Order parameter $\Delta\rho = \rho_l - \rho_g$ versus the reduced
  distance $x = \lambda/\lambda_c - 1$ on double-logarithmic axes; the
  red line is the two-dimensional Ising form $B\,x^{1/8}$ with the
  amplitude $B$ and $\lambda_c$ obtained from the joint
  fit. (b)~Coexistence diameter
  $\rho_{\mathrm{diam}} = (\rho_g + \rho_l)/2$ versus $\lambda$ with its
  linear fit $\rho_c + c\,(\lambda - \lambda_c)$; the star marks the
  critical point $(\rho_c, \lambda_c) = (0.112,\, 0.5824)$. Both panels
  use the same consensus coexistence data as
  Fig.~\ref{fig:binodal}(a).}
  \label{fig:gibbscrit:f15}
\end{figure*}

\begin{figure*}
  \includegraphics[width=\textwidth]{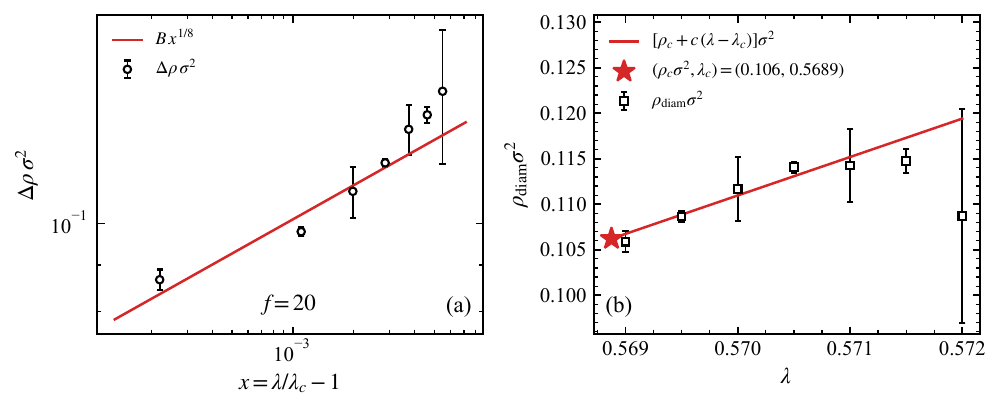}
  \caption{Same as Fig.~\ref{fig:gibbscrit:f15}, for $f = 20$; the
  critical point is $(\rho_c, \lambda_c) = (0.106,\, 0.5689)$. Both
  panels use the consensus coexistence data of
  Fig.~\ref{fig:binodal}(b).}
  \label{fig:gibbscrit:f20}
\end{figure*}

 The critical point is obtained from a joint,
error-weighted least-squares fit of Eq.~(\ref{eq:isingfit}), with
$\beta = 1/8$ imposed and free non-universal amplitude $B$, and the
linear diameter of Eq.~(\ref{eq:diameter}) to the consensus branches,
restricted to the near-critical windows
$0.5825 \le \lambda \le 0.585$ ($f = 15$) and
$0.569 \le \lambda \le 0.572$ ($f = 20$) where the asymptotic scaling
forms apply. The fit parameters are $\lambda_c$, $\rho_c$, $B$, and
the diameter slope $c$, with $\lambda_c$ constrained to lie below
every state point that exhibits phase separation. The resulting
scaling collapses are shown in Figs.~\ref{fig:gibbscrit:f15}
and~\ref{fig:gibbscrit:f20}. We emphasize that the sampled $\lambda$
windows do not determine the exponent independently; $\beta = 1/8$ is
a universality-class assumption, not a fitted result.

\section{Phase detection}
\label{app:solidliquid}

We detect the system's phase by analyzing orientational and translational order through their respective order parameters and correlation functions. The local 
bond-orientational order parameter is defined as
\begin{equation}
\psi_6(\mathbf{r}_j) = \frac{1}{N_b(j)} \sum_{m=1}^{N_b(j)} \exp \left( 6 \mathrm{i} \theta_{jm} \right),
\end{equation}
where the sum runs over the $N_b(j)$ nearest neighbors of particle $j$, and $\theta_{jm}$ is the angle between the bond connecting particles $j$ and $m$, and a fixed reference axis.

To characterize spatial correlations, we compute the orientational correlation function,
\begin{equation}
g_6(r) = \left\langle \psi_6(\mathbf{r}_i) \psi_6^*(\mathbf{r}_j) \right\rangle_{|\mathbf{r}_i - \mathbf{r}_j| = r},
\end{equation}
and the translational correlation function,
\begin{equation}
c_G(r) = \left\langle e^{\mathrm{i} \mathbf{G} \cdot (\mathbf{r}_i - \mathbf{r}_j)} 
\right\rangle_{|\mathbf{r}_i - \mathbf{r}_j| = r},
\end{equation}
where $\mathbf{G}$ is a reciprocal lattice vector of the triangular lattice.

\begin{figure*}
    \centering
    \includegraphics[width=\linewidth]{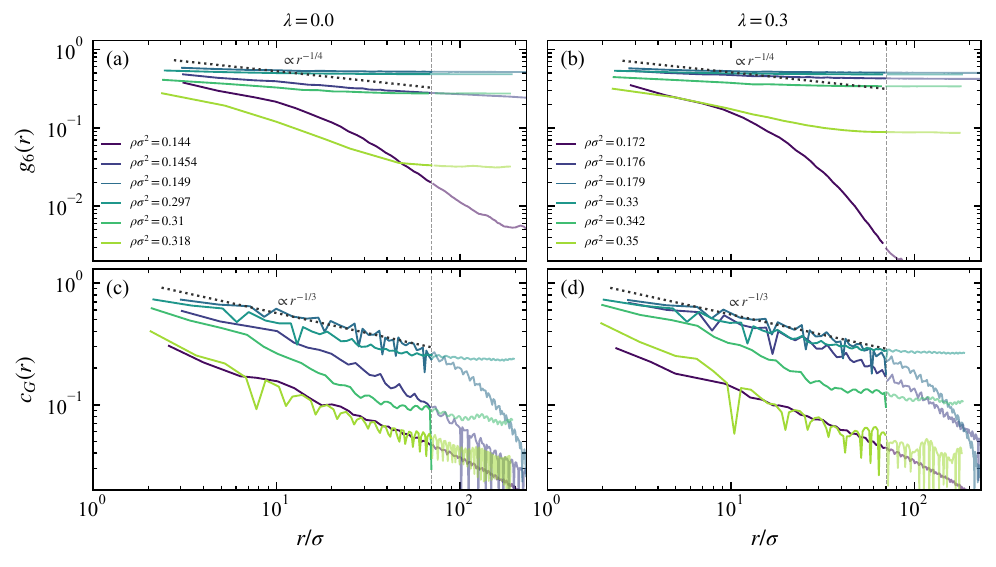}
    \caption{The phase evaluation for $\lambda =0.0$, and $\lambda=0.3$, $f = 15$. both for conventional and reentrant melting.}
    \label{fig:detect}
\end{figure*}

The phases are distinguished by the asymptotic behavior of these correlation functions:

\begin{itemize}
    \item \textbf{Solid phase:}
    \begin{align}
        c_G(r) &\sim r^{-\eta_G}, \\
        g_6(r) &\rightarrow \text{const.}
    \end{align}

    \item \textbf{Hexatic phase:}
    \begin{align}
        c_G(r) &\sim \exp(-r/\xi), \\
        g_6(r) &\sim r^{-\eta_6}.
    \end{align}

    \item \textbf{Liquid phase:}
    \begin{align}
        c_G(r) &\sim \exp(-r/\xi_T), \\
        g_6(r) &\sim \exp(-r/\xi_6),
    \end{align}
\end{itemize}
where $\xi$, $\xi_T$ and $\xi_6$ are correlation lengths
of the corresponding correlation functions. 

Figure \ref{fig:detect} shows the phase detection via the asymptotic behavior of the correlation functions. Our
resolution in density does not allow us to detect
the precise location of the hexatic phase, which 
occupies a very narrow region in the phase diagram.
The points in the plot present the solid transition points.


\begin{thebibliography}{56}\makeatletter
\providecommand \@ifxundefined [1]{\@ifx{#1\undefined}
}\providecommand \@ifnum [1]{\ifnum #1\expandafter \@firstoftwo
 \else \expandafter \@secondoftwo
 \fi
}\providecommand \@ifx [1]{\ifx #1\expandafter \@firstoftwo
 \else \expandafter \@secondoftwo
 \fi
}\providecommand \natexlab [1]{#1}\providecommand \enquote  [1]{``#1''}\providecommand \bibnamefont  [1]{#1}\providecommand \bibfnamefont [1]{#1}\providecommand \citenamefont [1]{#1}\providecommand \href@noop [0]{\@secondoftwo}\providecommand \href [0]{\begingroup \@sanitize@url \@href}\providecommand \@href[1]{\@@startlink{#1}\@@href}\providecommand \@@href[1]{\endgroup#1\@@endlink}\providecommand \@sanitize@url [0]{\catcode `\\12\catcode `\$12\catcode
  `\&12\catcode `\#12\catcode `\^12\catcode `\_12\catcode `\%12\relax}\providecommand \@@startlink[1]{}\providecommand \@@endlink[0]{}\providecommand \url  [0]{\begingroup\@sanitize@url \@url }\providecommand \@url [1]{\endgroup\@href {#1}{\urlprefix }}\providecommand \urlprefix  [0]{URL }\providecommand \Eprint [0]{\href }\providecommand \doibase [0]{https://doi.org/}\providecommand \selectlanguage [0]{\@gobble}\providecommand \bibinfo  [0]{\@secondoftwo}\providecommand \bibfield  [0]{\@secondoftwo}\providecommand \translation [1]{[#1]}\providecommand \BibitemOpen [0]{}\providecommand \bibitemStop [0]{}\providecommand \bibitemNoStop [0]{.\EOS\space}\providecommand \EOS [0]{\spacefactor3000\relax}\providecommand \BibitemShut  [1]{\csname bibitem#1\endcsname}\let\auto@bib@innerbib\@empty
\bibitem [{\citenamefont {Louis}(2002)}]{louis2002beware}\BibitemOpen
  \bibfield  {author} {\bibinfo {author} {\bibfnamefont {A.~A.}\ \bibnamefont
  {Louis}},\ }\bibfield  {title} {\bibinfo {title} {Beware of density dependent
  pair potentials},\ }\href {https://doi.org/10.1088/0953-8984/14/40/311}
  {\bibfield  {journal} {\bibinfo  {journal} {J. Phys.: Condens. Matter}\
  }\textbf {\bibinfo {volume} {14}},\ \bibinfo {pages} {9187} (\bibinfo {year}
  {2002})}\BibitemShut {NoStop}\bibitem [{\citenamefont {von Ferber}\ \emph {et~al.}(2000)\citenamefont {von
  Ferber}, \citenamefont {Jusufi}, \citenamefont {Likos}, \citenamefont
  {L{\"o}wen},\ and\ \citenamefont {Watzlawek}}]{vonferber2000triplet}\BibitemOpen
  \bibfield  {author} {\bibinfo {author} {\bibfnamefont {C.}~\bibnamefont {von
  Ferber}}, \bibinfo {author} {\bibfnamefont {A.}~\bibnamefont {Jusufi}},
  \bibinfo {author} {\bibfnamefont {C.~N.}\ \bibnamefont {Likos}}, \bibinfo
  {author} {\bibfnamefont {H.}~\bibnamefont {L{\"o}wen}},\ and\ \bibinfo
  {author} {\bibfnamefont {M.}~\bibnamefont {Watzlawek}},\ }\bibfield  {title}
  {\bibinfo {title} {Triplet interactions in star polymer solutions},\ }\href
  {https://doi.org/10.1007/s101890050013} {\bibfield  {journal} {\bibinfo
  {journal} {Eur. Phys. J. E}\ }\textbf {\bibinfo {volume} {2}},\ \bibinfo
  {pages} {311} (\bibinfo {year} {2000})}\BibitemShut {NoStop}\bibitem [{\citenamefont {Likos}(2001)}]{likos2001effective}\BibitemOpen
  \bibfield  {author} {\bibinfo {author} {\bibfnamefont {C.~N.}\ \bibnamefont
  {Likos}},\ }\bibfield  {title} {\bibinfo {title} {Effective interactions in
  soft condensed matter physics},\ }\href
  {https://doi.org/10.1016/S0370-1573(00)00141-1} {\bibfield  {journal}
  {\bibinfo  {journal} {Phys. Rep.}\ }\textbf {\bibinfo {volume} {348}},\
  \bibinfo {pages} {267} (\bibinfo {year} {2001})}\BibitemShut {NoStop}\bibitem [{\citenamefont {Likos}\ \emph {et~al.}(2001)\citenamefont {Likos},
  \citenamefont {Lang}, \citenamefont {Watzlawek},\ and\ \citenamefont
  {L{\"o}wen}}]{likos2001criterion}\BibitemOpen
  \bibfield  {author} {\bibinfo {author} {\bibfnamefont {C.~N.}\ \bibnamefont
  {Likos}}, \bibinfo {author} {\bibfnamefont {A.}~\bibnamefont {Lang}},
  \bibinfo {author} {\bibfnamefont {M.}~\bibnamefont {Watzlawek}},\ and\
  \bibinfo {author} {\bibfnamefont {H.}~\bibnamefont {L{\"o}wen}},\ }\bibfield
  {title} {\bibinfo {title} {Criterion for determining clustering versus
  reentrant melting behavior for bounded interaction potentials},\ }\href
  {https://doi.org/10.1103/PhysRevE.63.031206} {\bibfield  {journal} {\bibinfo
  {journal} {Phys. Rev. E}\ }\textbf {\bibinfo {volume} {63}},\ \bibinfo
  {pages} {031206} (\bibinfo {year} {2001})}\BibitemShut {NoStop}\bibitem [{\citenamefont {Prestipino}\ \emph {et~al.}(2005)\citenamefont
  {Prestipino}, \citenamefont {Saija},\ and\ \citenamefont
  {Giaquinta}}]{prestipino2005gcm}\BibitemOpen
  \bibfield  {author} {\bibinfo {author} {\bibfnamefont {S.}~\bibnamefont
  {Prestipino}}, \bibinfo {author} {\bibfnamefont {F.}~\bibnamefont {Saija}},\
  and\ \bibinfo {author} {\bibfnamefont {P.~V.}\ \bibnamefont {Giaquinta}},\
  }\bibfield  {title} {\bibinfo {title} {Phase diagram of the {G}aussian-core
  model},\ }\href {https://doi.org/10.1103/PhysRevE.71.050102} {\bibfield
  {journal} {\bibinfo  {journal} {Phys. Rev. E}\ }\textbf {\bibinfo {volume}
  {71}},\ \bibinfo {pages} {050102(R)} (\bibinfo {year} {2005})}\BibitemShut
  {NoStop}\bibitem [{\citenamefont {P{\`a}mies}\ \emph {et~al.}(2009)\citenamefont
  {P{\`a}mies}, \citenamefont {Cacciuto},\ and\ \citenamefont
  {Frenkel}}]{pamies2009hertz}\BibitemOpen
  \bibfield  {author} {\bibinfo {author} {\bibfnamefont {J.~C.}\ \bibnamefont
  {P{\`a}mies}}, \bibinfo {author} {\bibfnamefont {A.}~\bibnamefont
  {Cacciuto}},\ and\ \bibinfo {author} {\bibfnamefont {D.}~\bibnamefont
  {Frenkel}},\ }\bibfield  {title} {\bibinfo {title} {Phase diagram of
  {H}ertzian spheres},\ }\href {https://doi.org/10.1063/1.3186742} {\bibfield
  {journal} {\bibinfo  {journal} {J. Chem. Phys.}\ }\textbf {\bibinfo {volume}
  {131}},\ \bibinfo {pages} {044514} (\bibinfo {year} {2009})},\ \bibinfo
  {note} {erratum: \href{https://doi.org/10.1063/1.3251048}{J. Chem. Phys.
  \textbf{131}, 159903(E) (2009)}}\BibitemShut {NoStop}\bibitem [{\citenamefont {Hooton}(1955)}]{hooton1955anharmonicity}\BibitemOpen
  \bibfield  {author} {\bibinfo {author} {\bibfnamefont {D.~J.}\ \bibnamefont
  {Hooton}},\ }\bibfield  {title} {\bibinfo {title} {A new treatment of
  anharmonicity in lattice thermodynamics: {I}},\ }\href
  {https://doi.org/10.1080/14786440408520575} {\bibfield  {journal} {\bibinfo
  {journal} {Philos. Mag.}\ }\textbf {\bibinfo {volume} {46}},\ \bibinfo
  {pages} {422} (\bibinfo {year} {1955})}\BibitemShut {NoStop}\bibitem [{\citenamefont {Koehler}(1966)}]{koehler1966scp}\BibitemOpen
  \bibfield  {author} {\bibinfo {author} {\bibfnamefont {T.~R.}\ \bibnamefont
  {Koehler}},\ }\bibfield  {title} {\bibinfo {title} {Theory of the
  self-consistent harmonic approximation with application to solid neon},\
  }\href {https://doi.org/10.1103/PhysRevLett.17.89} {\bibfield  {journal}
  {\bibinfo  {journal} {Phys. Rev. Lett.}\ }\textbf {\bibinfo {volume} {17}},\
  \bibinfo {pages} {89} (\bibinfo {year} {1966})}\BibitemShut {NoStop}\bibitem [{\citenamefont {Werthamer}(1970)}]{werthamer1970scp}\BibitemOpen
  \bibfield  {author} {\bibinfo {author} {\bibfnamefont {N.~R.}\ \bibnamefont
  {Werthamer}},\ }\bibfield  {title} {\bibinfo {title} {Self-consistent phonon
  formulation of anharmonic lattice dynamics},\ }\href
  {https://doi.org/10.1103/PhysRevB.1.572} {\bibfield  {journal} {\bibinfo
  {journal} {Phys. Rev. B}\ }\textbf {\bibinfo {volume} {1}},\ \bibinfo {pages}
  {572} (\bibinfo {year} {1970})}\BibitemShut {NoStop}\bibitem [{\citenamefont {Lang}\ \emph {et~al.}(2000)\citenamefont {Lang},
  \citenamefont {Likos}, \citenamefont {Watzlawek},\ and\ \citenamefont
  {L{\"o}wen}}]{lang2000gcm}\BibitemOpen
  \bibfield  {author} {\bibinfo {author} {\bibfnamefont {A.}~\bibnamefont
  {Lang}}, \bibinfo {author} {\bibfnamefont {C.~N.}\ \bibnamefont {Likos}},
  \bibinfo {author} {\bibfnamefont {M.}~\bibnamefont {Watzlawek}},\ and\
  \bibinfo {author} {\bibfnamefont {H.}~\bibnamefont {L{\"o}wen}},\ }\bibfield
  {title} {\bibinfo {title} {Fluid and solid phases of the {G}aussian core
  model},\ }\href {https://doi.org/10.1088/0953-8984/12/24/302} {\bibfield
  {journal} {\bibinfo  {journal} {J. Phys.: Condens. Matter}\ }\textbf
  {\bibinfo {volume} {12}},\ \bibinfo {pages} {5087} (\bibinfo {year}
  {2000})}\BibitemShut {NoStop}\bibitem [{\citenamefont {Prestipino}(2012)}]{prestipino2012elastic}\BibitemOpen
  \bibfield  {author} {\bibinfo {author} {\bibfnamefont {S.}~\bibnamefont
  {Prestipino}},\ }\bibfield  {title} {\bibinfo {title} {On the accuracy of the
  melting curves drawn from modelling a solid as an elastic medium},\ }\href
  {https://doi.org/10.1088/0953-8984/24/3/035102} {\bibfield  {journal}
  {\bibinfo  {journal} {J. Phys.: Condens. Matter}\ }\textbf {\bibinfo {volume}
  {24}},\ \bibinfo {pages} {035102} (\bibinfo {year} {2012})}\BibitemShut
  {NoStop}\bibitem [{\citenamefont {Weeks}\ \emph {et~al.}(1971)\citenamefont {Weeks},
  \citenamefont {Chandler},\ and\ \citenamefont {Andersen}}]{wca1971}\BibitemOpen
  \bibfield  {author} {\bibinfo {author} {\bibfnamefont {J.~D.}\ \bibnamefont
  {Weeks}}, \bibinfo {author} {\bibfnamefont {D.}~\bibnamefont {Chandler}},\
  and\ \bibinfo {author} {\bibfnamefont {H.~C.}\ \bibnamefont {Andersen}},\
  }\bibfield  {title} {\bibinfo {title} {Role of repulsive forces in
  determining the equilibrium structure of simple liquids},\ }\href
  {https://doi.org/10.1063/1.1674820} {\bibfield  {journal} {\bibinfo
  {journal} {J. Chem. Phys.}\ }\textbf {\bibinfo {volume} {54}},\ \bibinfo
  {pages} {5237} (\bibinfo {year} {1971})}\BibitemShut {NoStop}\bibitem [{\citenamefont {Ramakrishnan}\ and\ \citenamefont
  {Yussouff}(1979)}]{ramakrishnan1979dft}\BibitemOpen
  \bibfield  {author} {\bibinfo {author} {\bibfnamefont {T.~V.}\ \bibnamefont
  {Ramakrishnan}}\ and\ \bibinfo {author} {\bibfnamefont {M.}~\bibnamefont
  {Yussouff}},\ }\bibfield  {title} {\bibinfo {title} {First-principles
  order-parameter theory of freezing},\ }\href
  {https://doi.org/10.1103/PhysRevB.19.2775} {\bibfield  {journal} {\bibinfo
  {journal} {Phys. Rev. B}\ }\textbf {\bibinfo {volume} {19}},\ \bibinfo
  {pages} {2775} (\bibinfo {year} {1979})}\BibitemShut {NoStop}\bibitem [{\citenamefont {Evans}(1979)}]{evans1979dft}\BibitemOpen
  \bibfield  {author} {\bibinfo {author} {\bibfnamefont {R.}~\bibnamefont
  {Evans}},\ }\bibfield  {title} {\bibinfo {title} {The nature of the
  liquid-vapour interface and other topics in the statistical mechanics of
  non-uniform, classical fluids},\ }\href
  {https://doi.org/10.1080/00018737900101365} {\bibfield  {journal} {\bibinfo
  {journal} {Adv. Phys.}\ }\textbf {\bibinfo {volume} {28}},\ \bibinfo {pages}
  {143} (\bibinfo {year} {1979})}\BibitemShut {NoStop}\bibitem [{\citenamefont {Likos}\ \emph {et~al.}(2007)\citenamefont {Likos},
  \citenamefont {Mladek}, \citenamefont {Gottwald},\ and\ \citenamefont
  {Kahl}}]{likos2007why}\BibitemOpen
  \bibfield  {author} {\bibinfo {author} {\bibfnamefont {C.~N.}\ \bibnamefont
  {Likos}}, \bibinfo {author} {\bibfnamefont {B.~M.}\ \bibnamefont {Mladek}},
  \bibinfo {author} {\bibfnamefont {D.}~\bibnamefont {Gottwald}},\ and\
  \bibinfo {author} {\bibfnamefont {G.}~\bibnamefont {Kahl}},\ }\bibfield
  {title} {\bibinfo {title} {Why do ultrasoft repulsive particles cluster and
  crystallize? {A}nalytical results from density-functional theory},\ }\href
  {https://doi.org/10.1063/1.2738064} {\bibfield  {journal} {\bibinfo
  {journal} {J. Chem. Phys.}\ }\textbf {\bibinfo {volume} {126}},\ \bibinfo
  {pages} {224502} (\bibinfo {year} {2007})}\BibitemShut {NoStop}\bibitem [{\citenamefont {van Teeffelen}\ \emph {et~al.}(2006)\citenamefont
  {van Teeffelen}, \citenamefont {Likos}, \citenamefont {Hoffmann},\ and\
  \citenamefont {L{\"o}wen}}]{vanteeffelen2006freezing}\BibitemOpen
  \bibfield  {author} {\bibinfo {author} {\bibfnamefont {S.}~\bibnamefont {van
  Teeffelen}}, \bibinfo {author} {\bibfnamefont {C.~N.}\ \bibnamefont {Likos}},
  \bibinfo {author} {\bibfnamefont {N.}~\bibnamefont {Hoffmann}},\ and\
  \bibinfo {author} {\bibfnamefont {H.}~\bibnamefont {L{\"o}wen}},\ }\bibfield
  {title} {\bibinfo {title} {Density functional theory of freezing for soft
  interactions in two dimensions},\ }\href
  {https://doi.org/10.1209/epl/i2006-10140-7} {\bibfield  {journal} {\bibinfo
  {journal} {Europhys. Lett.}\ }\textbf {\bibinfo {volume} {75}},\ \bibinfo
  {pages} {583} (\bibinfo {year} {2006})}\BibitemShut {NoStop}\bibitem [{\citenamefont {Parola}\ \emph {et~al.}(2008)\citenamefont {Parola},
  \citenamefont {Pini},\ and\ \citenamefont {Reatto}}]{parola2008smooth}\BibitemOpen
  \bibfield  {author} {\bibinfo {author} {\bibfnamefont {A.}~\bibnamefont
  {Parola}}, \bibinfo {author} {\bibfnamefont {D.}~\bibnamefont {Pini}},\ and\
  \bibinfo {author} {\bibfnamefont {L.}~\bibnamefont {Reatto}},\ }\bibfield
  {title} {\bibinfo {title} {Liquid-vapor transition from a microscopic theory:
  Beyond the {M}axwell construction},\ }\href
  {https://doi.org/10.1103/PhysRevLett.100.165704} {\bibfield  {journal}
  {\bibinfo  {journal} {Phys. Rev. Lett.}\ }\textbf {\bibinfo {volume} {100}},\
  \bibinfo {pages} {165704} (\bibinfo {year} {2008})}\BibitemShut {NoStop}\bibitem [{\citenamefont {Mendoza-Coto}\ \emph {et~al.}(2024)\citenamefont
  {Mendoza-Coto}, \citenamefont {Mattiello}, \citenamefont {Cenci},
  \citenamefont {Defenu},\ and\ \citenamefont {Nicolao}}]{MendozaCoto2024}\BibitemOpen
  \bibfield  {author} {\bibinfo {author} {\bibfnamefont {A.}~\bibnamefont
  {Mendoza-Coto}}, \bibinfo {author} {\bibfnamefont {V.}~\bibnamefont
  {Mattiello}}, \bibinfo {author} {\bibfnamefont {R.}~\bibnamefont {Cenci}},
  \bibinfo {author} {\bibfnamefont {N.}~\bibnamefont {Defenu}},\ and\ \bibinfo
  {author} {\bibfnamefont {L.}~\bibnamefont {Nicolao}},\ }\bibfield  {title}
  {\bibinfo {title} {Melting of the two-dimensional solid phase in the
  {G}aussian core model},\ }\href {https://doi.org/10.1103/PhysRevB.109.064101}
  {\bibfield  {journal} {\bibinfo  {journal} {Phys. Rev. B}\ }\textbf {\bibinfo
  {volume} {109}},\ \bibinfo {pages} {064101} (\bibinfo {year}
  {2024})}\BibitemShut {NoStop}\bibitem [{\citenamefont {Frenkel}\ and\ \citenamefont
  {Ladd}(1984)}]{frenkel1984ladd}\BibitemOpen
  \bibfield  {author} {\bibinfo {author} {\bibfnamefont {D.}~\bibnamefont
  {Frenkel}}\ and\ \bibinfo {author} {\bibfnamefont {A.~J.~C.}\ \bibnamefont
  {Ladd}},\ }\bibfield  {title} {\bibinfo {title} {New {M}onte {C}arlo method
  to compute the free energy of arbitrary solids. {A}pplication to the fcc and
  hcp phases of hard spheres},\ }\href {https://doi.org/10.1063/1.448024}
  {\bibfield  {journal} {\bibinfo  {journal} {J. Chem. Phys.}\ }\textbf
  {\bibinfo {volume} {81}},\ \bibinfo {pages} {3188} (\bibinfo {year}
  {1984})}\BibitemShut {NoStop}\bibitem [{\citenamefont {Vega}\ \emph {et~al.}(2008)\citenamefont {Vega},
  \citenamefont {Sanz}, \citenamefont {Abascal},\ and\ \citenamefont
  {Noya}}]{vega2008review}\BibitemOpen
  \bibfield  {author} {\bibinfo {author} {\bibfnamefont {C.}~\bibnamefont
  {Vega}}, \bibinfo {author} {\bibfnamefont {E.}~\bibnamefont {Sanz}}, \bibinfo
  {author} {\bibfnamefont {J.~L.~F.}\ \bibnamefont {Abascal}},\ and\ \bibinfo
  {author} {\bibfnamefont {E.~G.}\ \bibnamefont {Noya}},\ }\bibfield  {title}
  {\bibinfo {title} {Determination of phase diagrams via computer simulation:
  Methodology and applications to water, electrolytes and proteins},\ }\href
  {https://doi.org/10.1088/0953-8984/20/15/153101} {\bibfield  {journal}
  {\bibinfo  {journal} {J. Phys.: Condens. Matter}\ }\textbf {\bibinfo {volume}
  {20}},\ \bibinfo {pages} {153101} (\bibinfo {year} {2008})}\BibitemShut
  {NoStop}\bibitem [{\citenamefont {Likos}\ \emph
  {et~al.}(1998{\natexlab{a}})\citenamefont {Likos}, \citenamefont {L{\"o}wen},
  \citenamefont {Watzlawek}, \citenamefont {Abbas}, \citenamefont
  {Jucknischke}, \citenamefont {Allgaier},\ and\ \citenamefont
  {Richter}}]{Likos1998}\BibitemOpen
  \bibfield  {author} {\bibinfo {author} {\bibfnamefont {C.~N.}\ \bibnamefont
  {Likos}}, \bibinfo {author} {\bibfnamefont {H.}~\bibnamefont {L{\"o}wen}},
  \bibinfo {author} {\bibfnamefont {M.}~\bibnamefont {Watzlawek}}, \bibinfo
  {author} {\bibfnamefont {B.}~\bibnamefont {Abbas}}, \bibinfo {author}
  {\bibfnamefont {O.}~\bibnamefont {Jucknischke}}, \bibinfo {author}
  {\bibfnamefont {J.}~\bibnamefont {Allgaier}},\ and\ \bibinfo {author}
  {\bibfnamefont {D.}~\bibnamefont {Richter}},\ }\bibfield  {title} {\bibinfo
  {title} {Star polymers viewed as ultrasoft colloidal particles},\ }\href
  {https://doi.org/10.1103/PhysRevLett.80.4450} {\bibfield  {journal} {\bibinfo
   {journal} {Phys. Rev. Lett.}\ }\textbf {\bibinfo {volume} {80}},\ \bibinfo
  {pages} {4450} (\bibinfo {year} {1998}{\natexlab{a}})}\BibitemShut {NoStop}\bibitem [{\citenamefont {Jusufi}\ \emph {et~al.}(1999)\citenamefont {Jusufi},
  \citenamefont {Watzlawek},\ and\ \citenamefont
  {L{\"o}wen}}]{jusufi1999interaction}\BibitemOpen
  \bibfield  {author} {\bibinfo {author} {\bibfnamefont {A.}~\bibnamefont
  {Jusufi}}, \bibinfo {author} {\bibfnamefont {M.}~\bibnamefont {Watzlawek}},\
  and\ \bibinfo {author} {\bibfnamefont {H.}~\bibnamefont {L{\"o}wen}},\
  }\bibfield  {title} {\bibinfo {title} {Effective interaction between star
  polymers},\ }\href {https://doi.org/10.1021/ma981844u} {\bibfield  {journal}
  {\bibinfo  {journal} {Macromolecules}\ }\textbf {\bibinfo {volume} {32}},\
  \bibinfo {pages} {4470} (\bibinfo {year} {1999})}\BibitemShut {NoStop}\bibitem [{\citenamefont {Watzlawek}\ \emph {et~al.}(1999)\citenamefont
  {Watzlawek}, \citenamefont {Likos},\ and\ \citenamefont
  {L{\"o}wen}}]{Watzlawek1999}\BibitemOpen
  \bibfield  {author} {\bibinfo {author} {\bibfnamefont {M.}~\bibnamefont
  {Watzlawek}}, \bibinfo {author} {\bibfnamefont {C.~N.}\ \bibnamefont
  {Likos}},\ and\ \bibinfo {author} {\bibfnamefont {H.}~\bibnamefont
  {L{\"o}wen}},\ }\bibfield  {title} {\bibinfo {title} {Phase diagram of star
  polymer solutions},\ }\href {https://doi.org/10.1103/PhysRevLett.82.5289}
  {\bibfield  {journal} {\bibinfo  {journal} {Phys. Rev. Lett.}\ }\textbf
  {\bibinfo {volume} {82}},\ \bibinfo {pages} {5289} (\bibinfo {year}
  {1999})}\BibitemShut {NoStop}\bibitem [{\citenamefont {Farimani}\ and\ \citenamefont
  {Likos}(2025)}]{Farimani2025}\BibitemOpen
  \bibfield  {author} {\bibinfo {author} {\bibfnamefont {R.~A.}\ \bibnamefont
  {Farimani}}\ and\ \bibinfo {author} {\bibfnamefont {C.~N.}\ \bibnamefont
  {Likos}},\ }\bibfield  {title} {\bibinfo {title} {Coarse-graining of
  slit-confined star polymers in solvents of varying quality},\ }\href
  {https://doi.org/10.1021/acs.macromol.5c01343} {\bibfield  {journal}
  {\bibinfo  {journal} {Macromolecules}\ }\textbf {\bibinfo {volume} {58}},\
  \bibinfo {pages} {11827} (\bibinfo {year} {2025})}\BibitemShut {NoStop}\bibitem [{\citenamefont {Mayer}\ \emph {et~al.}(2009)\citenamefont {Mayer},
  \citenamefont {Sciortino}, \citenamefont {Likos}, \citenamefont {Tartaglia},
  \citenamefont {L{\"o}wen},\ and\ \citenamefont
  {Zaccarelli}}]{10.1021/ma801894x}\BibitemOpen
  \bibfield  {author} {\bibinfo {author} {\bibfnamefont {C.}~\bibnamefont
  {Mayer}}, \bibinfo {author} {\bibfnamefont {F.}~\bibnamefont {Sciortino}},
  \bibinfo {author} {\bibfnamefont {C.~N.}\ \bibnamefont {Likos}}, \bibinfo
  {author} {\bibfnamefont {P.}~\bibnamefont {Tartaglia}}, \bibinfo {author}
  {\bibfnamefont {H.}~\bibnamefont {L{\"o}wen}},\ and\ \bibinfo {author}
  {\bibfnamefont {E.}~\bibnamefont {Zaccarelli}},\ }\bibfield  {title}
  {\bibinfo {title} {Multiple glass transitions in star polymer mixtures:
  Insights from theory and simulations},\ }\href
  {https://doi.org/10.1021/ma801894x} {\bibfield  {journal} {\bibinfo
  {journal} {Macromolecules}\ }\textbf {\bibinfo {volume} {42}},\ \bibinfo
  {pages} {423} (\bibinfo {year} {2009})}\BibitemShut {NoStop}\bibitem [{\citenamefont {Egorov}\ \emph {et~al.}(2013)\citenamefont {Egorov},
  \citenamefont {Paturej}, \citenamefont {Likos},\ and\ \citenamefont
  {Milchev}}]{egorov2013adsorption}\BibitemOpen
  \bibfield  {author} {\bibinfo {author} {\bibfnamefont {S.~A.}\ \bibnamefont
  {Egorov}}, \bibinfo {author} {\bibfnamefont {J.}~\bibnamefont {Paturej}},
  \bibinfo {author} {\bibfnamefont {C.~N.}\ \bibnamefont {Likos}},\ and\
  \bibinfo {author} {\bibfnamefont {A.}~\bibnamefont {Milchev}},\ }\bibfield
  {title} {\bibinfo {title} {Controlling the interactions between soft colloids
  via surface adsorption},\ }\href {https://doi.org/10.1021/ma400188s}
  {\bibfield  {journal} {\bibinfo  {journal} {Macromolecules}\ }\textbf
  {\bibinfo {volume} {46}},\ \bibinfo {pages} {3648} (\bibinfo {year}
  {2013})}\BibitemShut {NoStop}\bibitem [{\citenamefont {Mermin}\ and\ \citenamefont
  {Wagner}(1966)}]{mermin1966wagner}\BibitemOpen
  \bibfield  {author} {\bibinfo {author} {\bibfnamefont {N.~D.}\ \bibnamefont
  {Mermin}}\ and\ \bibinfo {author} {\bibfnamefont {H.}~\bibnamefont
  {Wagner}},\ }\bibfield  {title} {\bibinfo {title} {Absence of ferromagnetism
  or antiferromagnetism in one- or two-dimensional isotropic {H}eisenberg
  models},\ }\href {https://doi.org/10.1103/PhysRevLett.17.1133} {\bibfield
  {journal} {\bibinfo  {journal} {Phys. Rev. Lett.}\ }\textbf {\bibinfo
  {volume} {17}},\ \bibinfo {pages} {1133} (\bibinfo {year}
  {1966})}\BibitemShut {NoStop}\bibitem [{\citenamefont {Hohenberg}(1967)}]{hohenberg1967}\BibitemOpen
  \bibfield  {author} {\bibinfo {author} {\bibfnamefont {P.~C.}\ \bibnamefont
  {Hohenberg}},\ }\bibfield  {title} {\bibinfo {title} {Existence of long-range
  order in one and two dimensions},\ }\href
  {https://doi.org/10.1103/PhysRev.158.383} {\bibfield  {journal} {\bibinfo
  {journal} {Phys. Rev.}\ }\textbf {\bibinfo {volume} {158}},\ \bibinfo {pages}
  {383} (\bibinfo {year} {1967})}\BibitemShut {NoStop}\bibitem [{\citenamefont {Mermin}(1968)}]{mermin1968crystal}\BibitemOpen
  \bibfield  {author} {\bibinfo {author} {\bibfnamefont {N.~D.}\ \bibnamefont
  {Mermin}},\ }\bibfield  {title} {\bibinfo {title} {Crystalline order in two
  dimensions},\ }\href {https://doi.org/10.1103/PhysRev.176.250} {\bibfield
  {journal} {\bibinfo  {journal} {Phys. Rev.}\ }\textbf {\bibinfo {volume}
  {176}},\ \bibinfo {pages} {250} (\bibinfo {year} {1968})}\BibitemShut
  {NoStop}\bibitem [{\citenamefont {Strandburg}(1988)}]{strandburg1988}\BibitemOpen
  \bibfield  {author} {\bibinfo {author} {\bibfnamefont {K.~J.}\ \bibnamefont
  {Strandburg}},\ }\bibfield  {title} {\bibinfo {title} {Two-dimensional
  melting},\ }\href {https://doi.org/10.1103/RevModPhys.60.161} {\bibfield
  {journal} {\bibinfo  {journal} {Rev. Mod. Phys.}\ }\textbf {\bibinfo {volume}
  {60}},\ \bibinfo {pages} {161} (\bibinfo {year} {1988})}\BibitemShut
  {NoStop}\bibitem [{\citenamefont {Chui}(1983)}]{chui1983}\BibitemOpen
  \bibfield  {author} {\bibinfo {author} {\bibfnamefont {S.~T.}\ \bibnamefont
  {Chui}},\ }\bibfield  {title} {\bibinfo {title} {Grain-boundary theory of
  melting in two dimensions},\ }\href {https://doi.org/10.1103/PhysRevB.28.178}
  {\bibfield  {journal} {\bibinfo  {journal} {Phys. Rev. B}\ }\textbf {\bibinfo
  {volume} {28}},\ \bibinfo {pages} {178} (\bibinfo {year} {1983})}\BibitemShut
  {NoStop}\bibitem [{\citenamefont {Bernard}\ and\ \citenamefont
  {Krauth}(2011)}]{bernard2011}\BibitemOpen
  \bibfield  {author} {\bibinfo {author} {\bibfnamefont {E.~P.}\ \bibnamefont
  {Bernard}}\ and\ \bibinfo {author} {\bibfnamefont {W.}~\bibnamefont
  {Krauth}},\ }\bibfield  {title} {\bibinfo {title} {Two-step melting in two
  dimensions: First-order liquid-hexatic transition},\ }\href
  {https://doi.org/10.1103/PhysRevLett.107.155704} {\bibfield  {journal}
  {\bibinfo  {journal} {Phys. Rev. Lett.}\ }\textbf {\bibinfo {volume} {107}},\
  \bibinfo {pages} {155704} (\bibinfo {year} {2011})}\BibitemShut {NoStop}\bibitem [{\citenamefont {Kapfer}\ and\ \citenamefont
  {Krauth}(2015)}]{kapfer2015}\BibitemOpen
  \bibfield  {author} {\bibinfo {author} {\bibfnamefont {S.~C.}\ \bibnamefont
  {Kapfer}}\ and\ \bibinfo {author} {\bibfnamefont {W.}~\bibnamefont
  {Krauth}},\ }\bibfield  {title} {\bibinfo {title} {Two-dimensional melting:
  From liquid-hexatic coexistence to continuous transitions},\ }\href
  {https://doi.org/10.1103/PhysRevLett.114.035702} {\bibfield  {journal}
  {\bibinfo  {journal} {Phys. Rev. Lett.}\ }\textbf {\bibinfo {volume} {114}},\
  \bibinfo {pages} {035702} (\bibinfo {year} {2015})}\BibitemShut {NoStop}\bibitem [{\citenamefont {Tsiok}\ \emph {et~al.}(2020)\citenamefont {Tsiok},
  \citenamefont {Gaiduk}, \citenamefont {Fomin},\ and\ \citenamefont
  {Ryzhov}}]{tsiok2020hertzian}\BibitemOpen
  \bibfield  {author} {\bibinfo {author} {\bibfnamefont {E.~N.}\ \bibnamefont
  {Tsiok}}, \bibinfo {author} {\bibfnamefont {E.~A.}\ \bibnamefont {Gaiduk}},
  \bibinfo {author} {\bibfnamefont {{\relax Yu}.~D.}\ \bibnamefont {Fomin}},\
  and\ \bibinfo {author} {\bibfnamefont {V.~N.}\ \bibnamefont {Ryzhov}},\
  }\bibfield  {title} {\bibinfo {title} {Melting scenarios of two-dimensional
  {H}ertzian spheres with a single triangular lattice},\ }\href
  {https://doi.org/10.1039/C9SM02262G} {\bibfield  {journal} {\bibinfo
  {journal} {Soft Matter}\ }\textbf {\bibinfo {volume} {16}},\ \bibinfo {pages}
  {3962} (\bibinfo {year} {2020})}\BibitemShut {NoStop}\bibitem [{\citenamefont {Thorneywork}\ \emph {et~al.}(2017)\citenamefont
  {Thorneywork}, \citenamefont {Abbott}, \citenamefont {Aarts},\ and\
  \citenamefont {Dullens}}]{thorneywork2017melting}\BibitemOpen
  \bibfield  {author} {\bibinfo {author} {\bibfnamefont {A.~L.}\ \bibnamefont
  {Thorneywork}}, \bibinfo {author} {\bibfnamefont {J.~L.}\ \bibnamefont
  {Abbott}}, \bibinfo {author} {\bibfnamefont {D.~G. A.~L.}\ \bibnamefont
  {Aarts}},\ and\ \bibinfo {author} {\bibfnamefont {R.~P.~A.}\ \bibnamefont
  {Dullens}},\ }\bibfield  {title} {\bibinfo {title} {Two-dimensional melting
  of colloidal hard spheres},\ }\href
  {https://doi.org/10.1103/PhysRevLett.118.158001} {\bibfield  {journal}
  {\bibinfo  {journal} {Phys. Rev. Lett.}\ }\textbf {\bibinfo {volume} {118}},\
  \bibinfo {pages} {158001} (\bibinfo {year} {2017})}\BibitemShut {NoStop}\bibitem [{\citenamefont {Kosterlitz}\ and\ \citenamefont
  {Thouless}(1973)}]{kosterlitz1973}\BibitemOpen
  \bibfield  {author} {\bibinfo {author} {\bibfnamefont {J.~M.}\ \bibnamefont
  {Kosterlitz}}\ and\ \bibinfo {author} {\bibfnamefont {D.~J.}\ \bibnamefont
  {Thouless}},\ }\bibfield  {title} {\bibinfo {title} {Ordering, metastability
  and phase transitions in two-dimensional systems},\ }\href
  {https://doi.org/10.1088/0022-3719/6/7/010} {\bibfield  {journal} {\bibinfo
  {journal} {J. Phys. C: Solid State Phys.}\ }\textbf {\bibinfo {volume} {6}},\
  \bibinfo {pages} {1181} (\bibinfo {year} {1973})}\BibitemShut {NoStop}\bibitem [{\citenamefont {Halperin}\ and\ \citenamefont
  {Nelson}(1978)}]{halperin1978}\BibitemOpen
  \bibfield  {author} {\bibinfo {author} {\bibfnamefont {B.~I.}\ \bibnamefont
  {Halperin}}\ and\ \bibinfo {author} {\bibfnamefont {D.~R.}\ \bibnamefont
  {Nelson}},\ }\bibfield  {title} {\bibinfo {title} {Theory of two-dimensional
  melting},\ }\href {https://doi.org/10.1103/PhysRevLett.41.121} {\bibfield
  {journal} {\bibinfo  {journal} {Phys. Rev. Lett.}\ }\textbf {\bibinfo
  {volume} {41}},\ \bibinfo {pages} {121} (\bibinfo {year} {1978})},\ \bibinfo
  {note} {erratum: \href{https://doi.org/10.1103/PhysRevLett.41.519}{Phys. Rev.
  Lett. \textbf{41}, 519(E) (1978)}}\BibitemShut {NoStop}\bibitem [{\citenamefont {Nelson}\ and\ \citenamefont
  {Halperin}(1979)}]{nelson1979}\BibitemOpen
  \bibfield  {author} {\bibinfo {author} {\bibfnamefont {D.~R.}\ \bibnamefont
  {Nelson}}\ and\ \bibinfo {author} {\bibfnamefont {B.~I.}\ \bibnamefont
  {Halperin}},\ }\bibfield  {title} {\bibinfo {title} {Dislocation-mediated
  melting in two dimensions},\ }\href
  {https://doi.org/10.1103/PhysRevB.19.2457} {\bibfield  {journal} {\bibinfo
  {journal} {Phys. Rev. B}\ }\textbf {\bibinfo {volume} {19}},\ \bibinfo
  {pages} {2457} (\bibinfo {year} {1979})}\BibitemShut {NoStop}\bibitem [{\citenamefont {Young}(1979)}]{young1979}\BibitemOpen
  \bibfield  {author} {\bibinfo {author} {\bibfnamefont {A.~P.}\ \bibnamefont
  {Young}},\ }\bibfield  {title} {\bibinfo {title} {Melting and the vector
  {C}oulomb gas in two dimensions},\ }\href
  {https://doi.org/10.1103/PhysRevB.19.1855} {\bibfield  {journal} {\bibinfo
  {journal} {Phys. Rev. B}\ }\textbf {\bibinfo {volume} {19}},\ \bibinfo
  {pages} {1855} (\bibinfo {year} {1979})}\BibitemShut {NoStop}\bibitem [{\citenamefont {Berezinskii}(1971)}]{berezinskii1971classical}\BibitemOpen
  \bibfield  {author} {\bibinfo {author} {\bibfnamefont {V.~L.}\ \bibnamefont
  {Berezinskii}},\ }\bibfield  {title} {\bibinfo {title} {Destruction of
  long-range order in one-dimensional and two-dimensional systems having a
  continuous symmetry group. {I}. {C}lassical systems},\ }\href
  {https://jetp.ras.ru/cgi-bin/dn/e_032_03_0493.pdf} {\bibfield  {journal}
  {\bibinfo  {journal} {Sov. Phys. JETP}\ }\textbf {\bibinfo {volume} {32}},\
  \bibinfo {pages} {493} (\bibinfo {year} {1971})}\BibitemShut {NoStop}\bibitem [{\citenamefont {Kosterlitz}(1974)}]{kosterlitz1974xy}\BibitemOpen
  \bibfield  {author} {\bibinfo {author} {\bibfnamefont {J.~M.}\ \bibnamefont
  {Kosterlitz}},\ }\bibfield  {title} {\bibinfo {title} {The critical
  properties of the two-dimensional {XY} model},\ }\href
  {https://doi.org/10.1088/0022-3719/7/6/005} {\bibfield  {journal} {\bibinfo
  {journal} {J. Phys. C: Solid State Phys.}\ }\textbf {\bibinfo {volume} {7}},\
  \bibinfo {pages} {1046} (\bibinfo {year} {1974})}\BibitemShut {NoStop}\bibitem [{\citenamefont {Zu}\ \emph {et~al.}(2016)\citenamefont {Zu},
  \citenamefont {Liu}, \citenamefont {Tong},\ and\ \citenamefont
  {Xu}}]{zu2016density}\BibitemOpen
  \bibfield  {author} {\bibinfo {author} {\bibfnamefont {M.}~\bibnamefont
  {Zu}}, \bibinfo {author} {\bibfnamefont {J.}~\bibnamefont {Liu}}, \bibinfo
  {author} {\bibfnamefont {H.}~\bibnamefont {Tong}},\ and\ \bibinfo {author}
  {\bibfnamefont {N.}~\bibnamefont {Xu}},\ }\bibfield  {title} {\bibinfo
  {title} {Density affects the nature of the hexatic-liquid transition in
  two-dimensional melting of soft-core systems},\ }\href
  {https://doi.org/10.1103/PhysRevLett.117.085702} {\bibfield  {journal}
  {\bibinfo  {journal} {Phys. Rev. Lett.}\ }\textbf {\bibinfo {volume} {117}},\
  \bibinfo {pages} {085702} (\bibinfo {year} {2016})}\BibitemShut {NoStop}\bibitem [{\citenamefont {Likos}\ \emph
  {et~al.}(1998{\natexlab{b}})\citenamefont {Likos}, \citenamefont {L{\"o}wen},
  \citenamefont {Poppe}, \citenamefont {Willner}, \citenamefont {Roovers},
  \citenamefont {Cubitt},\ and\ \citenamefont {Richter}}]{likos1998theta}\BibitemOpen
  \bibfield  {author} {\bibinfo {author} {\bibfnamefont {C.~N.}\ \bibnamefont
  {Likos}}, \bibinfo {author} {\bibfnamefont {H.}~\bibnamefont {L{\"o}wen}},
  \bibinfo {author} {\bibfnamefont {A.}~\bibnamefont {Poppe}}, \bibinfo
  {author} {\bibfnamefont {L.}~\bibnamefont {Willner}}, \bibinfo {author}
  {\bibfnamefont {J.}~\bibnamefont {Roovers}}, \bibinfo {author} {\bibfnamefont
  {B.}~\bibnamefont {Cubitt}},\ and\ \bibinfo {author} {\bibfnamefont
  {D.}~\bibnamefont {Richter}},\ }\bibfield  {title} {\bibinfo {title}
  {Ordering phenomena of star polymer solutions approaching the {$\Theta$}
  state},\ }\href {https://doi.org/10.1103/PhysRevE.58.6299} {\bibfield
  {journal} {\bibinfo  {journal} {Phys. Rev. E}\ }\textbf {\bibinfo {volume}
  {58}},\ \bibinfo {pages} {6299} (\bibinfo {year}
  {1998}{\natexlab{b}})}\BibitemShut {NoStop}\bibitem [{\citenamefont {Huissmann}\ \emph {et~al.}(2009)\citenamefont
  {Huissmann}, \citenamefont {Blaak},\ and\ \citenamefont
  {Likos}}]{huissmann2009solvents}\BibitemOpen
  \bibfield  {author} {\bibinfo {author} {\bibfnamefont {S.}~\bibnamefont
  {Huissmann}}, \bibinfo {author} {\bibfnamefont {R.}~\bibnamefont {Blaak}},\
  and\ \bibinfo {author} {\bibfnamefont {C.~N.}\ \bibnamefont {Likos}},\
  }\bibfield  {title} {\bibinfo {title} {Star polymers in solvents of varying
  quality},\ }\href {https://doi.org/10.1021/ma8023359} {\bibfield  {journal}
  {\bibinfo  {journal} {Macromolecules}\ }\textbf {\bibinfo {volume} {42}},\
  \bibinfo {pages} {2806} (\bibinfo {year} {2009})}\BibitemShut {NoStop}\bibitem [{\citenamefont {{Lo Verso}}\ \emph {et~al.}(2006)\citenamefont {{Lo
  Verso}}, \citenamefont {Likos},\ and\ \citenamefont
  {Reatto}}]{loverso2006attractions}\BibitemOpen
  \bibfield  {author} {\bibinfo {author} {\bibfnamefont {F.}~\bibnamefont {{Lo
  Verso}}}, \bibinfo {author} {\bibfnamefont {C.~N.}\ \bibnamefont {Likos}},\
  and\ \bibinfo {author} {\bibfnamefont {L.}~\bibnamefont {Reatto}},\
  }\bibfield  {title} {\bibinfo {title} {Star polymers with tunable
  attractions: Cluster formation, phase separation, reentrant
  crystallization},\ }in\ \href {https://doi.org/10.1007/3-540-32702-9_13}
  {\emph {\bibinfo {booktitle} {Smart Colloidal Materials}}},\ \bibinfo
  {series} {Progress in Colloid and Polymer Science}, Vol.\ \bibinfo {volume}
  {133},\ \bibinfo {editor} {edited by\ \bibinfo {editor} {\bibfnamefont
  {W.}~\bibnamefont {Richtering}}}\ (\bibinfo  {publisher} {Springer},\
  \bibinfo {address} {Berlin, Heidelberg},\ \bibinfo {year} {2006})\ pp.\
  \bibinfo {pages} {78--87}\BibitemShut {NoStop}\bibitem [{\citenamefont {Keim}\ \emph {et~al.}(2007)\citenamefont {Keim},
  \citenamefont {Maret},\ and\ \citenamefont {von
  Gr{\"u}nberg}}]{keim2007frank}\BibitemOpen
  \bibfield  {author} {\bibinfo {author} {\bibfnamefont {P.}~\bibnamefont
  {Keim}}, \bibinfo {author} {\bibfnamefont {G.}~\bibnamefont {Maret}},\ and\
  \bibinfo {author} {\bibfnamefont {H.~H.}\ \bibnamefont {von Gr{\"u}nberg}},\
  }\bibfield  {title} {\bibinfo {title} {{F}rank's constant in the hexatic
  phase},\ }\href {https://doi.org/10.1103/PhysRevE.75.031402} {\bibfield
  {journal} {\bibinfo  {journal} {Phys. Rev. E}\ }\textbf {\bibinfo {volume}
  {75}},\ \bibinfo {pages} {031402} (\bibinfo {year} {2007})}\BibitemShut
  {NoStop}\bibitem [{\citenamefont {Panagiotopoulos}(1987)}]{Panagiotopoulos1987}\BibitemOpen
  \bibfield  {author} {\bibinfo {author} {\bibfnamefont {A.~Z.}\ \bibnamefont
  {Panagiotopoulos}},\ }\bibfield  {title} {\bibinfo {title} {Direct
  determination of phase coexistence properties of fluids by {M}onte {C}arlo
  simulation in a new ensemble},\ }\href
  {https://doi.org/10.1080/00268978700101491} {\bibfield  {journal} {\bibinfo
  {journal} {Mol. Phys.}\ }\textbf {\bibinfo {volume} {61}},\ \bibinfo {pages}
  {813} (\bibinfo {year} {1987})}\BibitemShut {NoStop}\bibitem [{\citenamefont {Smit}\ \emph {et~al.}(1989)\citenamefont {Smit},
  \citenamefont {De~Smedt},\ and\ \citenamefont {Frenkel}}]{Smit1989}\BibitemOpen
  \bibfield  {author} {\bibinfo {author} {\bibfnamefont {B.}~\bibnamefont
  {Smit}}, \bibinfo {author} {\bibfnamefont {{\relax Ph}.}~\bibnamefont
  {De~Smedt}},\ and\ \bibinfo {author} {\bibfnamefont {D.}~\bibnamefont
  {Frenkel}},\ }\bibfield  {title} {\bibinfo {title} {Computer simulations in
  the {G}ibbs ensemble},\ }\href {https://doi.org/10.1080/00268978900102641}
  {\bibfield  {journal} {\bibinfo  {journal} {Mol. Phys.}\ }\textbf {\bibinfo
  {volume} {68}},\ \bibinfo {pages} {931} (\bibinfo {year} {1989})}\BibitemShut
  {NoStop}\bibitem [{\citenamefont {Sengupta}\ \emph {et~al.}(2000)\citenamefont
  {Sengupta}, \citenamefont {Nielaba},\ and\ \citenamefont
  {Binder}}]{sengupta2000elastic}\BibitemOpen
  \bibfield  {author} {\bibinfo {author} {\bibfnamefont {S.}~\bibnamefont
  {Sengupta}}, \bibinfo {author} {\bibfnamefont {P.}~\bibnamefont {Nielaba}},\
  and\ \bibinfo {author} {\bibfnamefont {K.}~\bibnamefont {Binder}},\
  }\bibfield  {title} {\bibinfo {title} {Elastic moduli, dislocation core
  energy, and melting of hard disks in two dimensions},\ }\href
  {https://doi.org/10.1103/PhysRevE.61.6294} {\bibfield  {journal} {\bibinfo
  {journal} {Phys. Rev. E}\ }\textbf {\bibinfo {volume} {61}},\ \bibinfo
  {pages} {6294} (\bibinfo {year} {2000})}\BibitemShut {NoStop}\bibitem [{\citenamefont {von Gr{\"u}nberg}\ \emph {et~al.}(2004)\citenamefont
  {von Gr{\"u}nberg}, \citenamefont {Keim}, \citenamefont {Zahn},\ and\
  \citenamefont {Maret}}]{vongrunberg2004elastic}\BibitemOpen
  \bibfield  {author} {\bibinfo {author} {\bibfnamefont {H.~H.}\ \bibnamefont
  {von Gr{\"u}nberg}}, \bibinfo {author} {\bibfnamefont {P.}~\bibnamefont
  {Keim}}, \bibinfo {author} {\bibfnamefont {K.}~\bibnamefont {Zahn}},\ and\
  \bibinfo {author} {\bibfnamefont {G.}~\bibnamefont {Maret}},\ }\bibfield
  {title} {\bibinfo {title} {Elastic behavior of a two-dimensional crystal near
  melting},\ }\href {https://doi.org/10.1103/PhysRevLett.93.255703} {\bibfield
  {journal} {\bibinfo  {journal} {Phys. Rev. Lett.}\ }\textbf {\bibinfo
  {volume} {93}},\ \bibinfo {pages} {255703} (\bibinfo {year}
  {2004})}\BibitemShut {NoStop}\bibitem [{\citenamefont {Daoud}\ and\ \citenamefont
  {Cotton}(1982)}]{Daoud1982}\BibitemOpen
  \bibfield  {author} {\bibinfo {author} {\bibfnamefont {M.}~\bibnamefont
  {Daoud}}\ and\ \bibinfo {author} {\bibfnamefont {J.~P.}\ \bibnamefont
  {Cotton}},\ }\bibfield  {title} {\bibinfo {title} {Star shaped polymers: A
  model for the conformation and its concentration dependence},\ }\href
  {https://doi.org/10.1051/jphys:01982004303053100} {\bibfield  {journal}
  {\bibinfo  {journal} {J. Phys. (Paris)}\ }\textbf {\bibinfo {volume} {43}},\
  \bibinfo {pages} {531} (\bibinfo {year} {1982})}\BibitemShut {NoStop}\bibitem [{\citenamefont {Wilding}\ and\ \citenamefont
  {Bruce}(1992)}]{wilding1992mixing}\BibitemOpen
  \bibfield  {author} {\bibinfo {author} {\bibfnamefont {N.~B.}\ \bibnamefont
  {Wilding}}\ and\ \bibinfo {author} {\bibfnamefont {A.~D.}\ \bibnamefont
  {Bruce}},\ }\bibfield  {title} {\bibinfo {title} {Density fluctuations and
  field mixing in the critical fluid},\ }\href
  {https://doi.org/10.1088/0953-8984/4/12/008} {\bibfield  {journal} {\bibinfo
  {journal} {J. Phys.: Condens. Matter}\ }\textbf {\bibinfo {volume} {4}},\
  \bibinfo {pages} {3087} (\bibinfo {year} {1992})}\BibitemShut {NoStop}\bibitem [{\citenamefont {Yang}(1952)}]{yang1952magnetization}\BibitemOpen
  \bibfield  {author} {\bibinfo {author} {\bibfnamefont {C.~N.}\ \bibnamefont
  {Yang}},\ }\bibfield  {title} {\bibinfo {title} {The spontaneous
  magnetization of a two-dimensional {I}sing model},\ }\href
  {https://doi.org/10.1103/PhysRev.85.808} {\bibfield  {journal} {\bibinfo
  {journal} {Phys. Rev.}\ }\textbf {\bibinfo {volume} {85}},\ \bibinfo {pages}
  {808} (\bibinfo {year} {1952})}\BibitemShut {NoStop}\bibitem [{\citenamefont {Smit}\ and\ \citenamefont
  {Frenkel}(1991)}]{Smit1991}\BibitemOpen
  \bibfield  {author} {\bibinfo {author} {\bibfnamefont {B.}~\bibnamefont
  {Smit}}\ and\ \bibinfo {author} {\bibfnamefont {D.}~\bibnamefont {Frenkel}},\
  }\bibfield  {title} {\bibinfo {title} {Vapor--liquid equilibria of the
  two-dimensional {L}ennard-{J}ones fluid(s)},\ }\href
  {https://doi.org/10.1063/1.460477} {\bibfield  {journal} {\bibinfo  {journal}
  {J. Chem. Phys.}\ }\textbf {\bibinfo {volume} {94}},\ \bibinfo {pages} {5663}
  (\bibinfo {year} {1991})}\BibitemShut {NoStop}\bibitem [{\citenamefont {Bos}\ \emph {et~al.}(2019)\citenamefont {Bos},
  \citenamefont {van~der Scheer}, \citenamefont {Ellenbroek},\ and\
  \citenamefont {Sprakel}}]{bos2019tails}\BibitemOpen
  \bibfield  {author} {\bibinfo {author} {\bibfnamefont {I.}~\bibnamefont
  {Bos}}, \bibinfo {author} {\bibfnamefont {P.}~\bibnamefont {van~der Scheer}},
  \bibinfo {author} {\bibfnamefont {W.~G.}\ \bibnamefont {Ellenbroek}},\ and\
  \bibinfo {author} {\bibfnamefont {J.}~\bibnamefont {Sprakel}},\ }\bibfield
  {title} {\bibinfo {title} {Two-dimensional crystals of star polymers: A tale
  of tails},\ }\href {https://doi.org/10.1039/C8SM02100G} {\bibfield  {journal}
  {\bibinfo  {journal} {Soft Matter}\ }\textbf {\bibinfo {volume} {15}},\
  \bibinfo {pages} {615} (\bibinfo {year} {2019})}\BibitemShut {NoStop}\bibitem [{\citenamefont {Zhu}\ \emph {et~al.}(2019)\citenamefont {Zhu},
  \citenamefont {Truskett},\ and\ \citenamefont {Bonnecaze}}]{zhu2019layer}\BibitemOpen
  \bibfield  {author} {\bibinfo {author} {\bibfnamefont {X.}~\bibnamefont
  {Zhu}}, \bibinfo {author} {\bibfnamefont {T.~M.}\ \bibnamefont {Truskett}},\
  and\ \bibinfo {author} {\bibfnamefont {R.~T.}\ \bibnamefont {Bonnecaze}},\
  }\bibfield  {title} {\bibinfo {title} {Phase diagram for two-dimensional
  layer of soft particles},\ }\href {https://doi.org/10.1039/C9SM00333A}
  {\bibfield  {journal} {\bibinfo  {journal} {Soft Matter}\ }\textbf {\bibinfo
  {volume} {15}},\ \bibinfo {pages} {4162} (\bibinfo {year}
  {2019})}\BibitemShut {NoStop}\end{thebibliography}
\end{document}